\documentclass[sigconf]{acmart}

\usepackage{balance,bm}
\usepackage{xcolor,colortbl}
\usepackage{url}
\usepackage{enumitem}
\usepackage{graphicx} 
\usepackage{textcomp}
\usepackage{subcaption}
\usepackage{mathtools}
\usepackage{algorithmic}
\usepackage[ruled]{algorithm2e}
\usepackage[T1]{fontenc}
\usepackage[utf8]{inputenc}
\usepackage{multirow, verbatim, xspace}
\usepackage{subcaption}
\usepackage{framed}

\definecolor{green}{rgb}{0.0, 0.65, 0.31}
\definecolor{bleudefrance}{rgb}{0.19, 0.55, 0.91}
\definecolor{ceruleanblue}{rgb}{0.16, 0.32, 0.75}
\definecolor{grey}{HTML}{969696}
\definecolor{lightgrey}{HTML}{d7d7d7}
\definecolor{greybackground}{HTML}{e9ecef}
\definecolor{violet}{HTML}{6a51a3}
\definecolor{lgreen}{HTML}{5ab4ac}
\definecolor{dgreen}{HTML}{005a32}
\definecolor{purple}{HTML}{ae017e}
\definecolor{orange}{HTML}{d95f0e}
\definecolor{mediumblue}{rgb}{0.0, 0.0, 0.8}
\definecolor{shadecolor}{gray}{0.94}

\newcommand{\scalebar}[1]{%
  \begingroup
  \leavevmode
  \dimen0=#1em\relax
  \dimen0=2\dimen0\relax

  \ifdim\dimen0<0pt
    \def\barcolor{orange}%
    \dimen2=-\dimen0\relax
    \raisebox{-3pt}{%
      \hbox{%
        \makebox[0pt][r]{\color{\barcolor}\rule{\dimen2}{4pt}}%
        \color{gray}\rule{0.4pt}{8pt}%
      }%
    }%
  \else
    \def\barcolor{bleudefrance}%
    \dimen2=\dimen0\relax
    \raisebox{-3pt}{%
      \hbox{%
        \color{gray}\rule{0.4pt}{8pt}%
        \color{\barcolor}\rule{\dimen2}{4pt}%
      }%
    }%
  \fi
  \endgroup
}
\colorlet{tablerowcolor}{gray!15} 
\colorlet{tablerowcolor2}{gray!12} 
\colorlet{tablerowcolor3}{gray!25} 

\newcommand{\rowcollight}{\rowcolor{tablerowcolor2}} %

\newif{\ifhidecomments}
\hidecommentsfalse 
\ifhidecomments
   \newcommand{\jiawei}[1]{}
   \newcommand{\munmun}[1]{}
   \newcommand{\ben}[1]{}
   \newcommand{\radium}[1]{}
   \newcommand{\mei}[1]{}
\else
   \newcommand{\jiawei}[1]{\textbf{\sffamily{\textcolor{lgreen}{[#1 -- Jiawei]}}}}
   \newcommand{\munmun}[1]{\textbf{\sffamily{\textcolor{purple}{[Munmun: #1]}}}} 
   \newcommand{\ben}[1]{\textbf{\sffamily{\textcolor{mediumblue}{[Ben: #1]}}}} 
   \newcommand{\radium}[1]{\textbf{\sffamily{\textcolor{orange}{[Radium: #1]}}}}  
   \newcommand{\mei}[1]{\textbf{\sffamily{\textcolor{olive}{[Mei: #1]}}}} 

\AtBeginDocument{%
  }

\copyrightyear{2026}
\acmYear{2026}
\setcopyright{cc}
\setcctype{by}
\acmConference[CHI '26]{Proceedings of the 2026 CHI Conference on Human Factors in Computing Systems}{April 13--17, 2026}{Barcelona, Spain}
\acmBooktitle{Proceedings of the 2026 CHI Conference on Human Factors in Computing Systems (CHI '26), April 13--17, 2026, Barcelona, Spain}
\acmPrice{}
\acmDOI{10.1145/3772318.3790316}
\acmISBN{979-8-4007-2278-3/2026/04}

\begin{document}

\title[How Large Language Models and Their Applications in Health are Represented Across Channels of Public Discourse]{AI as We Describe It: How Large Language Models and Their Applications in Health are Represented Across Channels of Public Discourse}

\author{Jiawei Zhou}
\affiliation{
  \institution{Georgia Institute of Technology}
  \city{Atlanta}
  \state{GA}
  \country{USA}
}
\email{j.zhou@gatech.edu}

\author{Lei Zhang}
\affiliation{
  \institution{Georgia Institute of Technology}
  \city{Atlanta}
  \state{GA}
  \country{USA}
}
\email{lzhang793@gatech.edu}

\author{Mei Li}
\affiliation{
  \institution{Georgia Institute of Technology}
  \city{Atlanta}
  \state{GA}
  \country{USA}
}
\email{mli783@gatech.edu}

\author{Benjamin D. Horne}
\affiliation{
  \institution{University of Tennessee Knoxville}
  \city{Knoxville}
  \state{TN}
  \country{USA}
}
\email{bhorne6@utk.edu}

\author{Munmun De Choudhury}
\affiliation{
  \institution{Georgia Institute of Technology}
  \city{Atlanta}
  \state{GA}
  \country{USA}
}
\email{munmund@gatech.edu}

\renewcommand{\shortauthors}{Jiawei Zhou et al.}

\begin{abstract}

Representation shapes public attitudes and behaviors. With the recent advances and rapid adoption of LLMs, the way these systems are introduced will negotiate societal expectations for their role in high-stakes domains like health. Yet it remains unclear whether current narratives present a balanced view. We analyzed five prominent discourse channels (news, research press, YouTube, TikTok, and Reddit) over a two-year period on lexical style, informational content, and symbolic representation. Discussions were generally positive and episodic, with positivity increasing over time. Risk communication was unthorough and often reduced to information quality incidents, while explanations of LLMs' generative nature were rare. Compared with professional outlets, TikTok and Reddit highlighted wellbeing applications and showed greater variations in tone and anthropomorphism but little attention to risks. We discuss implications for public discourse as a diagnostic tool in identifying literacy and governance gaps, and for communication and design strategies to support more informed LLM engagement.

\end{abstract}


  

\maketitle


\section{Introduction}

\begin{quote}
It is our use of a pile of bricks and mortar which makes it a `house'; and what we feel, think or say about it that makes a `house' a `home'. In part, we give things meaning by how we \textit{represent} them --- the words we use about them, the stories we tell about them, the images of them we produce, the emotions we associate with them, the ways we classify and conceptualize them, the values we place on them. 
--- Stuart Hall~\cite{hall1997representation}
\end{quote}

\noindent Large Language Models (LLMs) have rapidly captured public interest and have been embedded in many everyday information and communication systems like search engines~\cite{usnews2023ai, microsoft2023bing, google2023ai}. In the health domain, LLMs are being explored for tasks ranging from communication assistance~\cite{jo2023understanding, karinshak2023working} to decision support~\cite{thirunavukarasu2023large, tu2024towards}. 
However, evidence of these models' significant risks also has emerged, such as hallucinations and misinformation~\cite{xu2024hallucination, chang2024survey, zhou2023synthetic}, algorithmic bias~\cite{jin2024ask, moayeri2024worldbench, magu2025navigating}, and the potential to erode critical thinking and informed decision-making~\cite{lee2025impact, jakesch2023co, sharma2024generative}. As such, (inter)governmental organizations and unions like the World Health Organization and the European Union have urged careful evaluation and regulatory governance~\cite{world2024releases,eu_ai_act_2024}. 

A core reason these issues are so complex is that LLMs differ fundamentally from prior information and communication technologies. Unlike traditional tools designed primarily to retrieve or organize existing information, LLMs generate new content in a probabilistic manner~\cite{vaswani2017attention}, popularly known as ``next token predictors''~\cite{bender2021dangers}. This generative nature can create outputs that appear authoritative yet lack verifiable grounding or consistency~\cite{xu2024hallucination, zhou2023synthetic}. 
However, due to lagging public literacy and regulatory oversight, people's interactions are guided by pre-existing mental models and expectations shaped by prior technologies or even human relationships, including tendencies to anthropomorphize LLMs as intentional or emotionally aware~\cite{hill2025_psychosis}. As a result, users end up experimenting with this new technology for various reasons, often in ways that are no longer appropriate or sufficient for understanding and evaluating LLM outputs~\cite{zhang2024s, zhou2024risk}. 
When people perceive LLMs as human-like or even human replacements, these flawed understandings can lead to real-world harms, as seen by the emerging reports of LLM-infused systems being linked to criminal or suicidal cases~\cite{wilkins2024teen, Landymore2024psychologist, dupre2025kill} and contemporary discussions of ``AI psychosis'' as a new type or force of mental issues~\cite{Landymore2024psychologist}.

Given these technical shifts and the potential misalignment between how LLMs work and how people expect them to behave, public discourse's agenda-setting~\cite{mccombs1972agenda} function now plays an especial important role, as it ``may not be successful much of the time in telling people what to think, but it is stunningly successful in telling its readers what to think \textit{about}''~\cite{cohen2015press}. With the rapid and wide adoption of LLMs, people's understanding of these tools' capabilities and risks will determine if they can mindfully and meaningfully decide how to engage with them. Past work has shown that media not only shape public perceptions of the salience of societal issues~\cite{mccombs1972agenda}, but also influence behaviors and evaluations of emerging technologies and their potential impacts~\cite{das2014effect}. 

In other words, the way we introduce LLMs as an emerging technology will influence public perception and set expectations. Therefore, public discussions serve as both a proxy for public attitudes and interests in LLMs and as a force guiding future perceptions. However, less explored is whether current narratives offer a balanced view of both the potential and the limitations of LLMs. 

We respond to this gap by analyzing and describing the public communication of LLMs in health, a high-stakes domain that touches everyone's daily life. According to the agenda-setting theory~\cite{mccombs1972agenda, mccombs2018new}, there are two levels of agenda-setting in messages: the first-level agenda-setting is achieved through the \textit{``salience of issues''} that affects public exposure and attention, and the second-level agenda-setting is delivered through the \textit{``attributes of issues''} that influence public comprehension and sensation. Drawing on this, we examine both the \textit{salience} of LLM-related issues by analyzing which information elements are communicated in public discourse, as well as the \textit{attributes} of those issues by assessing the overall stylistic presentation of messages and the symbolic representation of message subjects (i.e., LLM entities). 
Specifically, this large-scale quantitative analysis characterizes how public discourse introduces the potential uses and risks of LLMs in the health domain across different channels by answering the following research questions:
\begin{itemize}
    \item [\textbf{RQ1.}] \textbf{[Lexical style]} What lexical tone and styles characterize public discourse on LLMs?
    \item [\textbf{RQ2.}] \textbf{[Informational content]} What information on narrative framing, health subdomains, and risk disclosure is communicated in public discourse on LLMs?
    \item [\textbf{RQ3.}] \textbf{[Symbolic representation]} To what extent are LLM entities anthropomorphically represented in public discourse?
\end{itemize}

To this end, we examined five prominent discourse channels between December 2022 and December 2024, based on the release date of ChatGPT (November 30, 2022)~\cite{ChatGPT_launch}. Each channel plays a unique role in shaping perceptions: news articles, research press releases, YouTube videos, TikTok videos, and Reddit posts. We collected 62,783 items using keyword combinations that required at least one health-related term and one LLM-related term, and reduced this to 21,773 items through LLM-assisted content filtering. We then studied the general trends and cross-platform differences in lexical style, informational content, and symbolic representation.
Our results demonstrate that public narratives about LLMs in health are generally positive and increasing so over time, and they are mostly framed episodically by focusing on isolated instances rather than broader societal or systemic implications. At the same time, we observe a lack of a thorough introduction to or an overview of LLM risks in public discourse. When risks are mentioned, they are largely framed as information quality concerns, with rare explanation of the generative nature that differentiates LLMs from traditional information sources. Further, differences exist across content creator types: layperson-driven platforms such as TikTok and Reddit emphasize consumer needs and mental health care while showing greater variations in emotional tone and anthropomorphism, yet are less likely to communicate risks. On the contrary, professionally-authored content such as news articles and research press releases focus more on clinical applications and are more likely to situate issues within broader societal or systemic contexts.

Overall, this work makes three key contributions. \textbf{(1)} We offer a large-scale and comprehensive examination of public discourse across five different communication channels. To the best of our knowledge, this is the first study to study discourse across both top-down (professionally created) and bottom-up (publicly generated) content on LLMs and their applications in health. 
\textbf{(2)} We categorize six health subdomains where LLMs can be applied, which can guide future research on understanding and evaluating LLM deployment in health contexts. 
\textbf{(3)} We provide empirical evidence on the current state of public discussions about LLMs and identify literacy and governance gaps. We discuss the implications for using discourse as a diagnostic tool to understand public perception and attention and for developing more effective communication approaches and design strategies to support user agency and knowledge for meaningful and mindful engagement with LLMs.


\section{Related Work}

\subsection{Media Effects on Public Perceptions of Emerging Technology}
Media, ranging from mass media to social media, has long been recognized for its role in both reflecting and shaping public perceptions and attitudes~\cite{mccombs1972agenda, sharma2017analyzing, cascini2022social}. For instance, media exposure has been shown to drive public engagement in national conversations~\cite{king2017news}, influence views on the importance and priorities of regulatory issues~\cite{mccombs1972agenda, kellstedt2003mass, feezell2018agenda}, and can both inform \cite{dutton2011next, halpern2013social} and misguide \cite{ecker2024misinformation} the public. 

With regards to emerging technology, media plays a key role in communicating the sociotechnical risks and benefits~\cite{boholm1998comparative, marks2007mass, weaver2009searching}. Pubic discourse coverage can affect which aspects of a technology are emphasized, which in turn shapes how the public evaluates its value and impact. For example, public discourse can emphasize or de-emphasize who is responsible for benefits and harms of a technology, specific use cases of the technology, or the need for regulation~\cite{weaver2009searching}. As a result, evidence from studies and opinion polls suggests that public attitudes towards emerging technology can ``mirror the news media's stance''~\cite{marks2007mass, bauer2002controversial}. Specifically, prior work has studied media coverage of emerging technologies such as nanotechnology~\cite{anderson2005framing} and AI~\cite{allaham2025informing, ittefaq2025global, ouchchy2020ai}, including relevant risks~\cite{weaver2009searching, allaham2025informing} and technological issues~\cite{das2018breaking}. Broadly, these works have demonstrated that media tends to embrace a positive outlook about new technology, focusing more on the benefits than the potential risks~\cite{anderson2005framing, wilkinson2007uncertainty}, although with some technologies this trend has changed over time~\cite{weaver2009searching}. In the case of AI, \citet{ouchchy2020ai} found that media portrayals of AI ethics issues were shallow and largely focused on practical implications.

\subsection{Public Attitudes Toward LLMs and Generative AI}

Generative AI and specifically large language models (LLMs) are unique as their use in consumer-facing products has been quickly widespread~\cite{Reuters2024openai}, letting the public experiment with the technology themselves ahead of adequate public literacy and regulatory oversight. This public experience, alongside the rapid technological development, likely changes both how influential the public discourse is in perception formation and how the public discourse covers the technology. As such, public narratives surrounding LLMs and LLM-infused systems may be more participatory and reflective of personal experiences than those surrounding prior emerging technologies like nanotechnology. 

Recent work has begun to explore public perceptions of AI-generated content~\cite{lloyd2025ai, kadoma2025generative} and perceptions of AI itself~\cite{moreira2025hall, cheng2025tools}. However, overall findings remain mixed, and public attitudes were found to vary across socio-demographic factors~\cite{ravvselj2025higher, moreira2025hall}. Some work points to skepticism toward AI-generated content. \citet{kadoma2025generative} studied the perceptual harms of generative AI and found that people tended to associate AI writing with lower quality. Other studies have also shown negative perceptions about AI-generated content. For example, \citet{wang2025ai} found that people felt negatively about AI-generated art on TikTok due to both qualities of the content itself (e.g., hyper-realistic or unsettling qualities of the paintings) and principled stands against AI in art (e.g., ``unfairly copying or appropriating the work of human artists''). On the other hand, other research indicates favorable perceptions. \citet{park2024ai} discovered that people perceived AI Instagram accounts as more attractive and the quality of their content as similar as influencer-created content. More broadly, \citet{cheng2025tools} found that Americans generally viewed AI as warm and competent, and this perception has significantly increased over time. 

Overall, current work suggests that public attitudes toward AI are neither uniformly positive nor negative, but instead context-dependent. Evidence shows that both AI usage and attitudes are mixed and vary across industries~\cite{vasiljeva2021artificial} and across populations~\cite{moreira2025hall, ravvselj2025higher}. For instance, a global study of college students found that students generally felt positive about using ChatGPT, but its use varied across socio-demographic and geographic factors \cite{ravvselj2025higher}. 

While these studies offer valuable insights into how the public perceives LLMs and LLM-generated content, they largely focus on specific platforms and contexts. There remains a gap in holistically understanding how LLMs are represented across diverse discourse channels, including both professional-led and public-driven sources. Our study builds on prior scholarship by offering a large-scale examination of both top-down (professionally created) and bottom-up (publicly generated) content.

\subsection{LLMs and Generative AI in Health and Healthcare}

Besides the fact that people are experiencing LLMs before adequate understanding and oversight, LLMs are also distinctive due to the breadth of proposed use cases, ranging from general-purpose ones such as LLM-infused search engines~\cite{ico_search} to specific ones like LLM-infused health record systems~\cite{Landi2024epic}. In this study, we focus on LLMs within the high-stakes domain of health that influences people's daily life. In health, LLMs have been shown to have potential in supporting a variety of tasks such as clinical documentation~\cite{thirunavukarasu2023large, zhang2024generalist}, medical question answering and reasoning~\cite{nori2023capabilities, zhang2024generalist, ayers2023comparing}, clinical decision support~\cite{thirunavukarasu2023large, tu2024towards}, therapeutic conversations~\cite{wang2021evaluation, sharma2024facilitating, kim2024mindfuldiary}, and public health interventions~\cite{jo2023understanding, jo2024understanding, karinshak2023working}. In practice, LLMs have already been implemented in existing systems or workflows, including Electronic Health Records (EHR) systems ~\cite{Landi2024epic}, virtual agents for dealing with domestic violence~\cite{mccall2023ai}, and public health interventions~\cite{jo2023understanding}. Further, public-facing LLM chatbots like ChatGPT are being used for seeking health information \cite{yun2025online}.

Evidence so far suggests that perceptions of generative AI in healthcare have been mixed but skew positive, similar to perceptions of generative AI in broader contexts. A study by \citet{heinrichs2025physicians} showed that AI in medicine is viewed positively, with physicians reporting high AI enthusiasm and low AI skepticism. On the other hand, though a randomized experiment, \citet{yang2025peer} demonstrated that practicing clinicians perceived physicians using generative AI as a primary decision-making tool as having ``significantly lower in clinical skill.'' Still, the same study supported general enthusiasm about generative AI's use in healthcare. Others have similarly suggested that patients have a positive view of generative AI in health. For instance, through a large global survey, \citet{busch2025multinational} found that the majority of patients held positive views about AI in healthcare. However, once again, perceptions varied widely across socio-demographic groups and technical literacy.

\subsection{Risks of LLMs and Generative AI}

Despite the largely positive view of generative AI among the public, researchers have raised concerns about the limitations and risks of adopting LLMs in the real world. General issues include representativeness of training data~\cite{wornow2023shaky}, privacy and security~\cite{zhang2024s}, language and geographic disparities~\cite{jin2024ask, moayeri2024worldbench}, the generation of inaccurate, biased, or toxic content~\cite{liang2022holistic, chang2024survey, zhou2023synthetic}, and the meaningfulness of evaluation metrics~\cite{wornow2023shaky}.  Because of these limitations, questions remain regarding LLMs' ability to meet clinical standards~\cite{harrer2023attention}, avoid race-based medical misconceptions~\cite{omiye2023large}, attend to emotional needs~\cite{wang2021evaluation, sharma2024facilitating}, or handle high-stakes or high-intensity cases~\cite{sharma2024facilitating, kim2024mindfuldiary}. Such pitfalls are not simply technical challenges; they may contribute to serious real-world harms, as seen by the emerging reports of LLM-infused systems being linked to criminal or suicidal cases~\cite{wilkins2024teen, Landymore2024psychologist, dupre2025kill}.

Research has shown that people tend to struggle to recognize technological changes and instead rely on perceived reliability~\cite{mohanty2025lies}. In the case of LLMs, work has suggested that people tend to hold diverse and sometimes flawed mental models of LLM systems, which are relevant to AI knowledge. Evidence has showed that people tended to have overly simplified and erroneous mental models of data handling and model mechanisms, which limited risk awareness and judgment~\cite{zhang2024s}. People without basic AI literacy skills struggle more in critically engaging with LLMs~\cite{prabhudesai2025here}. Regarding these gaps in understanding and attitudes, research suggests that they are relevant to AI literacy and societal framing. Studies have identified polarization in AI attitudes linked to differences in technological knowledge, with misalignment between AI influencers and the U.S. general population~\cite{moreira2025hall} and especially strong polarization in tech-centric communities~\cite{qi2024excitements}. 

What is worse, gaps in risk perceptions and AI literacy can be amplified by society's imbalanced representations of LLMs' capabilities and limitations, potentially leading to societal hype. \citet{drogt2024ethical} criticized claims of AI outperforming medical practitioners, as much of the supporting evidence is not empirically convincing or transparently reported. Global news media on AI risks also had a skewed focus on societal, legal and rights-related risks, which may leave the public with an incomplete understanding of potential harms~\cite{allaham2025informing}. However, despite these findings, less is known about how LLMs are framed and symbolically represented across channels of discourse, leaving open questions about how conceptualizations of a novel technology may shape public understanding and expectations. We respond to this gap by considering how societal issues and AI's identity are framed in public discussions.


\begin{figure*}[t]
\centering
\includegraphics[width=\textwidth]{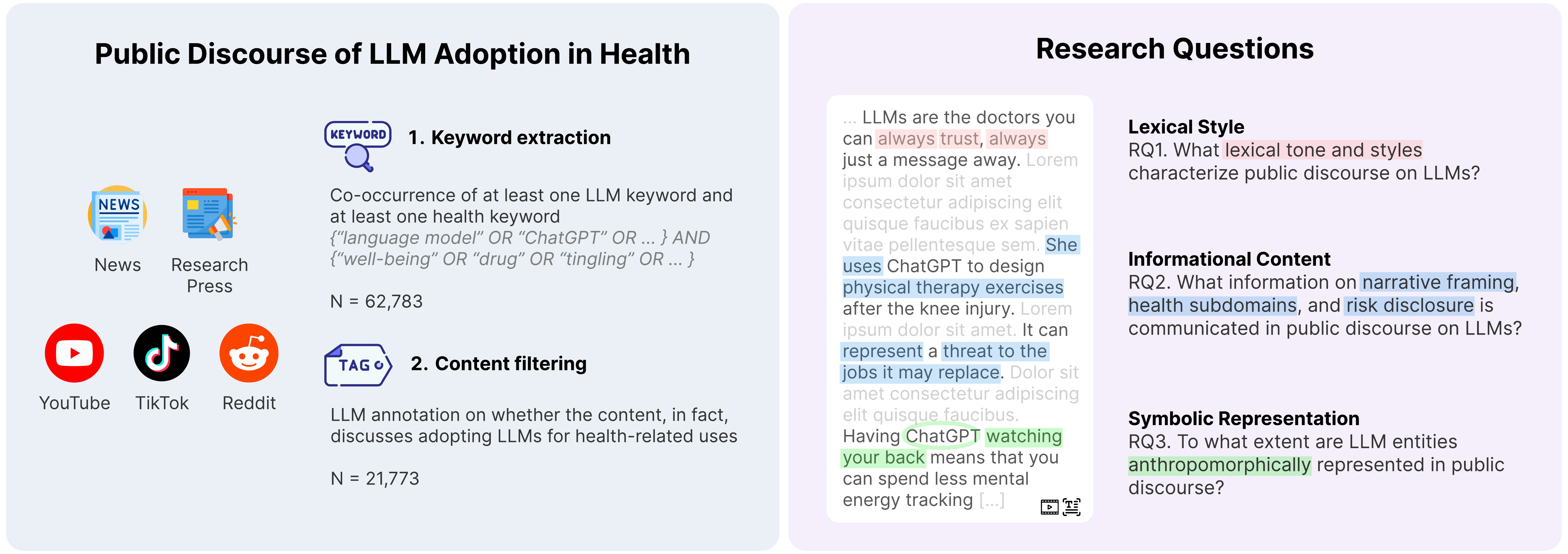}
\caption{Overview of our study investigating how public discourse introduces LLMs and their applications in the health domain. We collected pubic narratives from five prominent discourse channels, including news, research press, YouTube, TikTok, and Reddit, between December 2022 and December 2024. Drawing on agenda-setting theory, we examined three core dimensions of discourse:
(1) Lexical Style: the overall presentation style of discourse through emotional tone and writing formality;
(2) Informational Content: how messages frame LLMs' implications, risks, mechanisms, and potential across health domains;
(3) Symbolic Representation: the level of anthropomorphic representation of LLM entities.}
\label{fig:overview}
\Description{Overview of our study investigating how public discourse introduces LLMs and their applications in the health domain. We collected public narratives from five prominent discourse channels—news, research press releases, YouTube, TikTok, and Reddit—between December 2022 and December 2024. These narratives were first extracted based on the co-occurrence of at least one LLM-related keyword and at least one health-related keyword, and then further filtered using LLM annotation to determine whether the content actually discusses the adoption of LLMs for health-related uses. Drawing on agenda-setting theory, we then examined three core dimensions of discourse: (1) Lexical Style: the overall presentation style of discourse through emotional tone and writing formality; (2) Informational Content: how messages frame LLMs' implications, risks, mechanisms, and potential across health domains; (3) Symbolic Representation: the level of anthropomorphic representation of LLM entities.}
\end{figure*}

\section{Method}

This study presents a large-scale, multi-platform analysis of how large language models (LLMs) and their applications in health are represented in public discourse. As shown in Fig.~\ref{fig:overview}, we examined five prominent discourse channels: news articles, research press releases, YouTube videos, TikTok videos, and Reddit posts --- each playing a distinct role in shaping public understanding and attention. Using health- and AI-related keywords, we first collected data across these platforms and applied an additional round of LLM-assisted filtering to ensure content relevance to LLM adoption in health fields. Our analysis focuses on three key dimensions:
\textbf{(1)} \textit{Lexical style}, which captures overall presentation style through emotional tone and writing style; \textbf{(2)} \textit{Informational content}, which analyzes the mentioning of specific information surrounding framing, health subdomains, risk and mechanism disclosure; and \textbf{(3)} \textit{Symbolic representation}, which examines the framing of the AI's identity through the level of anthropomorphism attributed to LLM entities. To assess differences across channels, we employed Kruskal–Wallis H tests to compare distributions across groups for each dimension, followed by Dunn's post hoc tests with Bonferroni correction for pairwise comparisons.


\subsection{Data}

We studied five distinct media sources: news articles, research press releases, YouTube videos, TikTok videos, and Reddit posts. Each source plays a unique role in shaping perceptions. 
To elaborate, news coverage by journalists aims to make complex issues accessible to general audiences but may oversimplify nuances or overemphasize aspects that are more likely to resonate with the public. 
In contrast, research press releases are official statements produced by scientific organizations, such as universities and medical centers, to share research findings with the general public. They represent more in-depth and expert-led coverage but tend to be less accessible and have a narrower focus. YouTube, as a platform for video content, serves as a space for both professional and user-generated content, where discussions about LLMs can vary widely in terms of depth and perspective. 
TikTok, with its younger user base \cite{montag2021psychology}, presents a more informal, quick-hit style of content, and it may prioritize entertainment over depth and objectiveness. 
Lastly, Reddit's community-driven structure and pseudo-anonymous design foster conversations among individuals with shared interests and encourage more focused discussions about emerging technologies and personal experiences with those technologies~\cite{qi2024excitements}.

\vspace{0.5em}
\noindent\textbf{\textit{Ethics Statement:}} As we used publicly accessible data without any direct interactions with individuals, this work did not require institutional review board approval. However, we took careful attention in managing and presenting our data and followed the ethical practices in social media research by removing deleted or removed content and paraphrasing social media content quoted as examples.

\subsubsection{Data Source and Collection:}
\label{section:data_collection}
We collected public narratives in English over a two-year period, from December 2022 to December 2024, based on the release date of ChatGPT (November 30, 2022)~\cite{ChatGPT_launch}. To curate related content discussing LLM for health, we used a combination of health- and LLM-related keywords with an `OR' operator within a category and an `AND' operator across, ensuring each item contains at least one LLM-relevant keyword (e.g., \textit{``language model''} or \textit{``ChatGPT''}) and at least one health and symptom term (e.g., \textit{``well-being''}, \textit{``drug''}, or \textit{``tingling''}). The complete keyword list can be found in Appendix~\ref{appendix:keywords}. These keyword was adapted from search strategies in existing review studies and WHO-listed health topic words~\cite{liu2023summary, who_healthTopics, patel2025systematic}. For instance, \citet{patel2025systematic} used \textit{`Generative AI/GenAI/Large Language Models''} as the search terms for systematic reviews.
After several rounds of sanity checks by the authors, keywords were refined by removing ones that over-introduced false positives. For example, `gemini' was replaced with `google gemini' to avoid confusion with the word `gemini' in astronomy. 
In total, we collected 62,783 items of articles, posts, and videos. The data size for each source is presented in Table~\ref{table:data}. Specifically, data from each discourse channel was collected as follows: 

\begin{itemize}[leftmargin=1.1em]
    \item \textbf{News:} News articles were extracted from extended versions of the \texttt{NELA-GT}~\cite{gruppi2021nela} and \texttt{NELA-Local}~\cite{horne2022nela} datasets using the same collection pipelines, which include articles published by U.S. national and local news media outlets. We limited media outlets to those deemed reliable mainstream sources or mixed reliable sources, based on the Media Bias Fact Check (MBFC)~\cite{bmfc}, an independent resource maintained by researchers and journalists to evaluate the bias and credibility of media sources. The goal is to exclude fringe news without standard editorial oversight.
    
    \item \textbf{Research Press:} Research press release data was collected from the \texttt{EurekAlert}~\cite{EurekAlert} news-release distribution operated by the American Association for the Advancement of Science. EurekAlert hosts press releases produced by organizations that engage in all disciplines of scientific research, such as universities, medical centers, and government agencies. 
    
    \item \textbf{YouTube and TikTok:} For video data, we first identified relevant videos in English using \textit{YouTube Data API} and \textit{TikTok Research API}, and then collected and transcribed identified videos for subsequent analyses. We decided not to rely on the transcripts available through \textit{YouTube Transcript API} and \textit{TikTok Research API} due to inconsistent transcription quality on YouTube API and the inconsistent availability of built-in voice-to-text features on TikTok API. Instead, we referred to prior work~\cite{nguyen2025supporters} and implemented a video transcription pipeline using \texttt{yt-dlp} to download videos, \texttt{ffmpeg} to extract audio, and \texttt{whisper}~\cite{radford2023robust} to transcribe the audio content. This approach allowed us to collect transcriptions for the majority of videos, except for 693 videos consisting of pure music or no audio, 158 videos that were inaccessible due to privacy restrictions, 169 videos with encoding issue that prevented audio extraction, and 155 videos with gibberish content. Additionally, an internal TikTok API error 
    \footnote{Similar issue has been reported online: e.g., https://stackoverflow.com/questions/78731820/how-to-resolve-error-code-500-for-tiktok-research-api; https://www.reddit.com/r/CompSocial/comments/1d8mdz5/tiktok\_api/} 
    prevented data collection starting from August 18, 2024. We had submitted a support ticket to TikTok, but the API internal issue was not resolved at the time of writing and revision.
    
    \item \textbf{Reddit:} Reddit data was extracted from the widely used PushShift dataset~\cite{baumgartner2020pushshift}, with deleted and removed content excluded from analyses. We only included the original submissions, not the replies, to ensure sufficient context for analyses.
\end{itemize}


\subsubsection{Content Filtering:}

While the above crafted keyword combination ensures the extracted content mentions words or phrases about both LLMs and health, it is possible that such expressions appear in unrelated contexts, such as ``people \textit{abusing} \textit{ChatGPT} for writing their essays.'' Therefore, we conducted a round of targeted filtering to ensure that the content, in fact, discussed adopting LLM for health-related uses. We used \texttt{GPT-4o-mini}~\cite{hurst2024gpt} with few-shot examples to assist this additional round of content filtering. We referred to previous work~\cite{mittal2025online} and set the temperature as 0.0, provided four examples (3 positive examples and 1 negative example), and emphasized that annotations are based solely on the literal writing of the text. The complete prompt is provided in the supplement. 
For cases where the content exceeded the 128k token limit, the articles were split into several chunks, and the annotation results were combined. After this targeted filtering, the final dataset was reduced from 62,783 items to 21,773 items. Table~\ref{table:data} provides a summary of item numbers and content length statistics across the five sources.
To assess the reliability of utilizing an LLM for filtering articles, we sampled 100 articles across five data sources for human annotation. The model performed well on this task with an F1-macro score of 0.9 and Cohen's $\kappa$ of 0.7947.

\begin{table}[t]
\centering
\sffamily
\small
\caption{Descriptive summary of the number of items and word length across the five channels of public discourse.}
\label{table:data}
\resizebox{\columnwidth}{!}{%
\begin{tabular}{lrrrrr}
    \rowcollight & \multicolumn{2}{c}{\textbf{Items}} & \multicolumn{3}{c}{\textbf{Length} \textit{(\# words)}}\\
    \rowcollight \textbf{Source} & (pre filter) & (post filter) & Mean & Median & Std dev\\
    \toprule
    News            & 6 550      & 2 254     &  1 158.74    &  898      & 1 247.01\\
    Research Press  & 867       & 634       &  805.96      &  742      & 411.14\\
    YouTube         & 6 632     & 3 559     &   5 451.65    &  3 228    & 8 958.01\\
    TikTok          & 9 521     & 3 272     &   276.07      &  228      & 234.29\\
    Reddit          & 39 213    & 12 054   &  510.81        &  297      & 697.64\\
    \cline{2-3}
     & 62 783 & 21 773 & & & \\
    \bottomrule
\end{tabular}
}
\end{table}



\subsection{Lexical Style}

To understand the overall presentation style in different discourse channels portraying LLMs, we analyzed both emotional tone and writing style, as these summary patterns would better capture the overall communication styles than other direct word-category counts. We analyzed emotional tone to assess the pattern of affective expressions, to understand how different discourse channels portray LLMs in emotionally charged or neutral ways. For this, we utilized the Linguistic Inquiry and Word Count (LIWC)~\cite{Tausczik2010TheMethods} program, which is a validated psycholinguistic lexicon and has been widely used in linguistic and social science research to study emotional and cognitive processes in text~\cite{Jiang2018LinguisticMedia, zhou2023synthetic}.
\textit{Emotional tone} is quantified on a continuous scale, with values close to 0 indicating a highly negative sentiment, 50 representing a neutral sentiment, and values approaching 100 corresponding to a highly positive sentiment. \textit{Analytic style} (or categorical-dynamic index, CDI~\cite{Pennebaker2014WhenEssays}) captures expressions of abstract thinking and cognitive complexity, which are associated with formal and logical thinking patterns. \textit{Clout style} reflects the prevalence of self-focused expressions that indicate relative social status, confidence, and leadership, which tends to capture perceived authority, confidence, and credibility and is commonly studied in health communication and discourse~\cite{calle2024towards, karinshak2023working}. Lastly, \textit{authentic style} indicates the level of spontaneous and non-regulated language and is commonly used in self-disclosures.

\subsection{Informational Content}

Following our analysis of writing style, we analyzed three features about the informational content: what \textit{health subdomains} are mentioned, how each discourse channel \textit{frames} LLMs in health, and if and how \textit{risks} and LLM mechanisms are disclosed in the content.


\subsubsection{Framing Types and Dimensions}
We analyzed how content is framed by examining 1) \textit{framing type} --- whether the narrative is framed episodically or thematically, and 2) \textit{framing dimension} --- what broader policy-related themes are framed as relevant. These two framing typologies were proposed by \citet{iyengar1994anyone} and \citet{boydstun2013identifying} respectively. \textit{Framing type} focuses on how the story is told, distinguishing between episodic and thematic approaches. Specifically, episodic framing presents concrete information about specific people, places, or events, often focusing on individual stories or case studies. On the other hand, thematic framing provides a more generalized perspective on issues, placing stories in broader political and social contexts. \textit{Framing dimension} identifies frame cues across 14 policy issues, categorizing the content within broader societal and political contexts such as health, technology, or societal impact. To examine these framings, we used labels for both narrative and issue framing from a dataset of 4500 manually annotated data~\cite{mendelsohn2021modeling}. Then, for each type of frame, we built a multilabel \texttt{RoBERTa} classification model, replicating prior work~\cite{mendelsohn2021modeling} to classify the frames in all items across the five discourse channels. This model has been used in prior work to study gender violence in news~\cite{mittal2024news}.
Using both the original definitions by \citet{boydstun2013identifying} and \citet{iyengar1994anyone} and the comprehensive codebook by \citet{mendelsohn2021modeling}, we annotated 100 items to evaluate the applicability of this model on our dataset and found the model performed with an F1-macro average score of 0.7761 (Precision: 0.8, Recall: 0.7536).

\subsubsection{Risk and Generative Nature Disclosure.}
We considered whether the content discussed the potential risks of LLMs and the differences in mechanisms from traditional information technology. To assess these aspects, we utilized a risk taxonomy focused on LLM adoption for public health~\cite{zhou2024risk}.
This typology was selected for its relevance to the health domain and its greater applicability compared to other existing taxonomies. Based on focus groups with health professionals and individuals with experience in seeking health information, this work identified distinctive characteristics of LLMs that differentiate them from traditional information sources and summarized potential negative consequences of adopting LLMs for health informational needs~\cite{zhou2024risk}. The taxonomy identifies four risk dimensions: (1) risks to individuals: how LLMs may negatively influence individuals' actions, decisions, and overall well-being, (2) risks to human-centered care: how LLMs may undermine the relations and systems of healthcare and social support network, (3) risks to the information ecosystem: how LLMs may affect the people, technologies, and norms that shape the way information is produced, shared, and evaluated, (4) and risks to technology accountability: how LLMs may challenge existing mechanisms of oversight, regulation, and security in technologies.  
In addition to the risk categories, we also explored whether the content explained the \textit{generative nature} of LLMs (i.e., mentioning of probabilistic nature and limitations such as no real understanding of content) compared to non-generative AI or information retrieval-based systems~\cite{vaswani2017attention, bender2021dangers, zhou2024risk}. 

We utilized \texttt{GPT-4o-mini}~\cite{hurst2024gpt} with few-shot examples to annotate the presence or absence of each risk category and its generative explanation. This approach allowed us to efficiently scale the annotation process across the corpus. 
Consistent with previous task settings described above, we set the temperature to 0.0 to ensure deterministic output and provided four examples to guide the model in making accurate classifications.
To assess the reliability of utilizing LLM for categorization, we randomly sampled 100 articles across five data sources for human annotation, where risk disclosure had an F1-macro score of 0.74 (Precision: 0.72, Recall: 0.79), and generative nature explanation had an F1-macro score of 0.85 (Precision: 0.95, Recall: 0.80). To further examine potential model-specific bias, we conducted an additional cross-model validation using \texttt{gemini-2.5} on a random subset of 500 items. \texttt{gpt-4o-mini} showed slightly higher F1-macro scores with good inter-model agreements, indicating stable annotation behavior on our data. We report both the model–human evaluation (n = 100) and model–model agreement (n = 500) statistics in Appendix~\ref{appendix:model_agreement}.

\subsubsection{Health Subdomain Prevalence}
The potential applications of LLMs in health span a wide range, from providing healthy lifestyle advice to supporting mental health and professional medical training. Therefore, to understand the distribution and prevalence of specific health subdomains mentioned in the content, we combined human annotation with the use of \texttt{GPT-4o-mini}~\cite{hurst2024gpt} to scale up the categorization across the entire corpus. Specifically, we began by conducting conventional content analysis~\cite{hsieh2005three} to summarize the health subdomains referenced in the content, based on a random sample of 200 items that were equally distributed across the five discourse channels. Two researchers independently reviewed all the sampled items to establish a general understanding of the data. Afterwards, they used an inductive and iterative approach to develop new codes or categorize each instance into existing codes. In the meantime, the codes were continuously revisited for potential modification and refinement, either by breaking them into sub-categories or by grouping them into broader categories. By the end of the analysis, we determined that the categorization had reached saturation without emerging meaningful subdomains. 
This subdomain categorization was reviewed by two experts: a practicing medical doctor with research experience and a researcher specializing in public health communication. This expert consultation helped us validate and improve the clarity of all categories. 

The final categorization comprised 20 subcategories, organized into six broad groups: clinical decision support, workflow optimization, professional training, research and data analysis, consumer health support, and healthcare system support. The complete set of subcategories is presented in the Results Sect~\ref{result:content}. To scale up the categorization across the entire corpus for prevalence analysis, we employed \texttt{GPT-4o-mini}~\cite{hurst2024gpt} using the same temperature settings of 0.0 as previous tasks to ensure deterministic output. The prompt is provided in the supplement.
When the content and prompt together exceeded the token limit, the content was split into multiple chunks, and the annotation results were combined for final reporting. 
To assess the reliability of utilizing LLM for categorization, we sampled 100 articles across five data sources for human annotation. The model performed fine on this task with an F1-macro score of 0.792 (Precision: 0.745, Recall: 0.888). Same as previous tasks, we conducted an additional cross-model validation using \texttt{gemini-2.5} on a random subset of 500 items. \texttt{gpt-4o-mini} showed slightly higher F1-macro scores than \texttt{gemini-2.5} with high inter-model agreement, indicating stable annotation behavior on our data. We report both the model–human evaluation (n = 100) and model–model agreement (n = 500) statistics in Appendix~\ref{appendix:model_agreement}.

\subsection{Symbolic Representation}

Lastly, to examine the symbolic representation of LLMs in public discourse, we measured the level of implicit anthropomorphism in expressions containing LLM entities. We used a masked language model to capture the degree of anthropomorphism associated with LLMs in a given sentence. This approach proposed by \citet{cheng2024anthroscore} utilizes a masking technique to compare how much an entity is implicitly framed as human versus non-human, by computing the log of the ratio between the model's output probabilities for replacing the mask with human pronouns versus non-human pronouns. Specifically, for each instance in our corpus, when a sentence mentioned any LLM entity, the full sentence was extracted, with the LLM entity replaced with a \texttt{<MASK>} token. In addition to the list of LLM entities provided in the original work~\cite{cheng2024anthroscore}, we supplemented the model with our LLM-related keywords used in the content filtering step. 
An anthropomorphism probability was then calculated for each identified entity in a masked sentence (Equation~\ref{equation:anthro}, where $Anthro$ captures the degree of anthropomorphism for entity $x$ in sentence $s$), which was then used to calculate the average anthropomorphism score across all identified masked sentences within a given item. With this approach, an anthropomorphism score of 0 means that the LLM entity is equally likely to be stated as either human or non-human, and positive scores mean that the LLM entity is more likely to be implicitly framed as human in the sentence.

\vspace{-0.5em}
\begin{equation}
\label{equation:anthro}
Anthro(s_x) = \log \frac{P_{\text{Human}}(s_x)}{P_{\text{Non-Human}}(s_x)} .
\end{equation}


\section{Results}

\begin{figure*}[t]
\small
\centering
\includegraphics[width=0.65\textwidth]{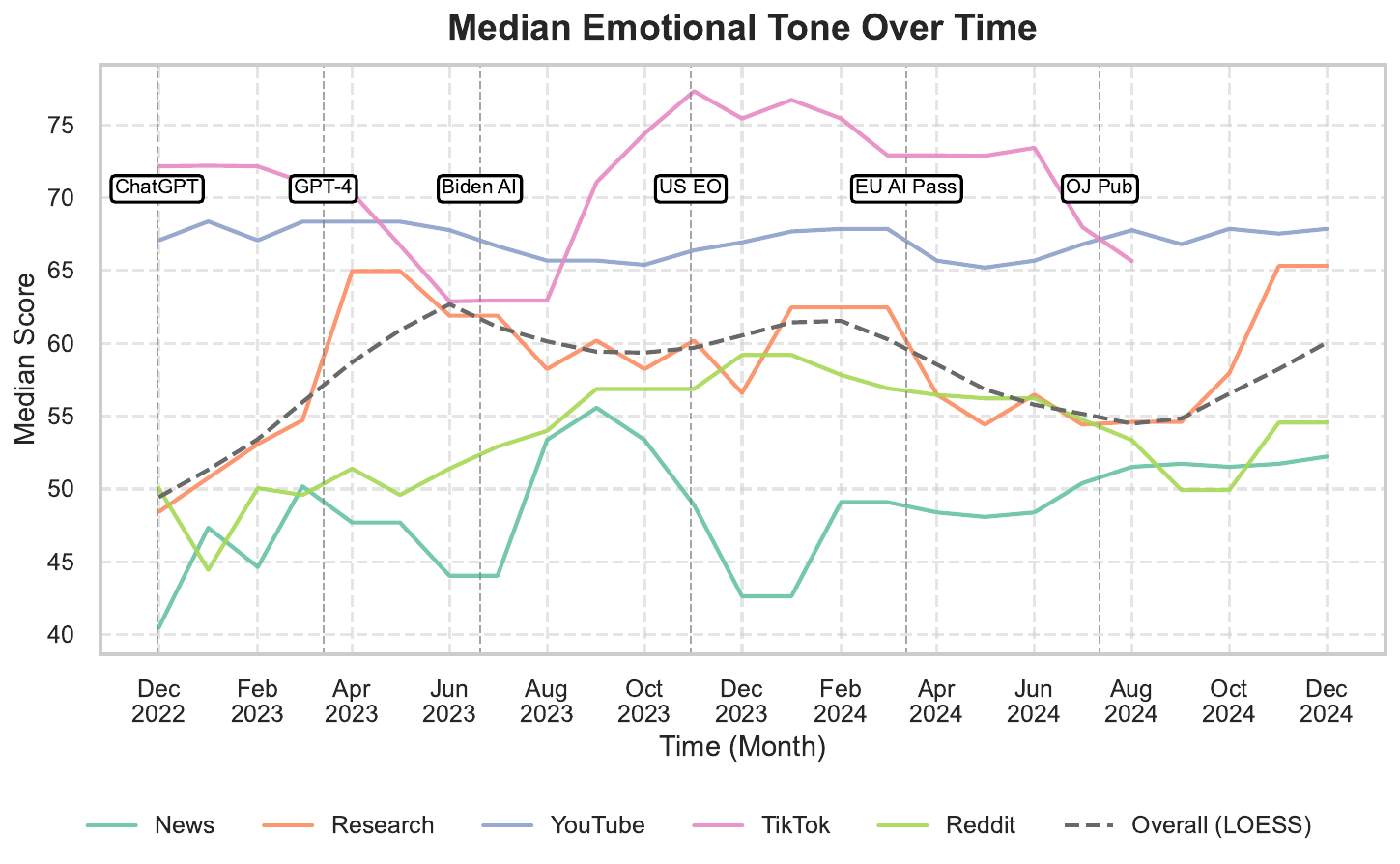}
\caption{Trends in median emotional tone, where public discussions of LLMs in health contexts were largely positive (except for news) and became increasingly positive over time. TikTok data ended in August'24 due to an API internal error (see Sec.~\ref{section:data_collection}). LOESS (Locally Estimated Scatterplot Smoothing) was applied to capture overall trends across data sources of varying size.}
\label{fig:emotion_trend}
\Description{A multi-line time series graph shows the trends in median emotional tone over time from Dec 2022 to Dec 2024. Multiple colored lines represent different content sources: News, Research, YouTube, TikTok, Reddit, and an overall smoothed trend (LOESS). Overall, public discussions of LLMs in health contexts were largely positive (except for news) and became increasingly positive over time. The overall smoothed trend suggests a gradual increase from early 2023 into mid-2023, followed by variability with slight upward movement toward late 2024. TikTok and YouTube showed the most optimistic tone, while News had an average neutral tone.}
\end{figure*}

\subsection{Lexical Style}

\begin{shaded}
\noindent\textbf{Summary:} (1) Public discussions on LLM for health were generally positive, and this positivity has gradually increased over time. TikTok and YouTube showed the most optimistic tone, whereas the News had an average neutral tone. (2) Writing styles mirrored these patterns: News and Research Press had highly analytical styles, consistent with professional norms. In contrast, social media platforms were less analytical and TikTok and YouTube scored higher on clout, suggesting that content creators presented themselves as confident and persuasive. Across all channels, authentic and spontaneous communication was low, particularly in professional outlets.
\end{shaded}

\subsubsection{Emotional Tone.}

Overall, the emotional tone of public discussions on LLM for health was generally positive, with most discourse channels (except for News) showing a positive emotional tone on average. This suggests that the portrayal of LLMs in health contexts was largely optimistic. In addition, as shown in Fig.~\ref{fig:emotion_trend}, we found the median emotional tone in public discourse on LLMs had increased over time despite some fluctuations. However, there was notable variation across channels. TikTok and YouTube content tended to exhibit the most positive tone, with higher median and mean values. For example, one video says \textit{``ChatGPT designed a \textbf{unique} \textbf{healthy} recipe that tasted \textbf{great} [...] I had one minor tweak, but overall \textbf{great} tool, \textbf{great} time-saver, very \textbf{exciting} new technology.''}
Dunn's post hoc test further demonstrated that YouTube and TikTok content was significantly more positive than other sources. In contrast, News and Reddit discussions showed a more neutral tone, with Reddit in particular having the widest range of emotional attitudes. Dunn's post hoc test indicated that News was significantly more negative than other sources. This could be explained by the nature of these platforms: Reddit's community-driven nature, which encourages diverse opinions and facilitates focused discussions, while news articles are more likely to reflect journalistic norms that emphasize balance between highlighting opportunities and addressing risks. The technological orientation of Reddit's community may also have contributed to this high variation, as \citet{qi2024excitements} has shown that tech-centric communities exhibit greater polarization.

\subsubsection{Writing Style.}

\begin{figure*}
\small
\centering
\includegraphics[width=0.75\textwidth]{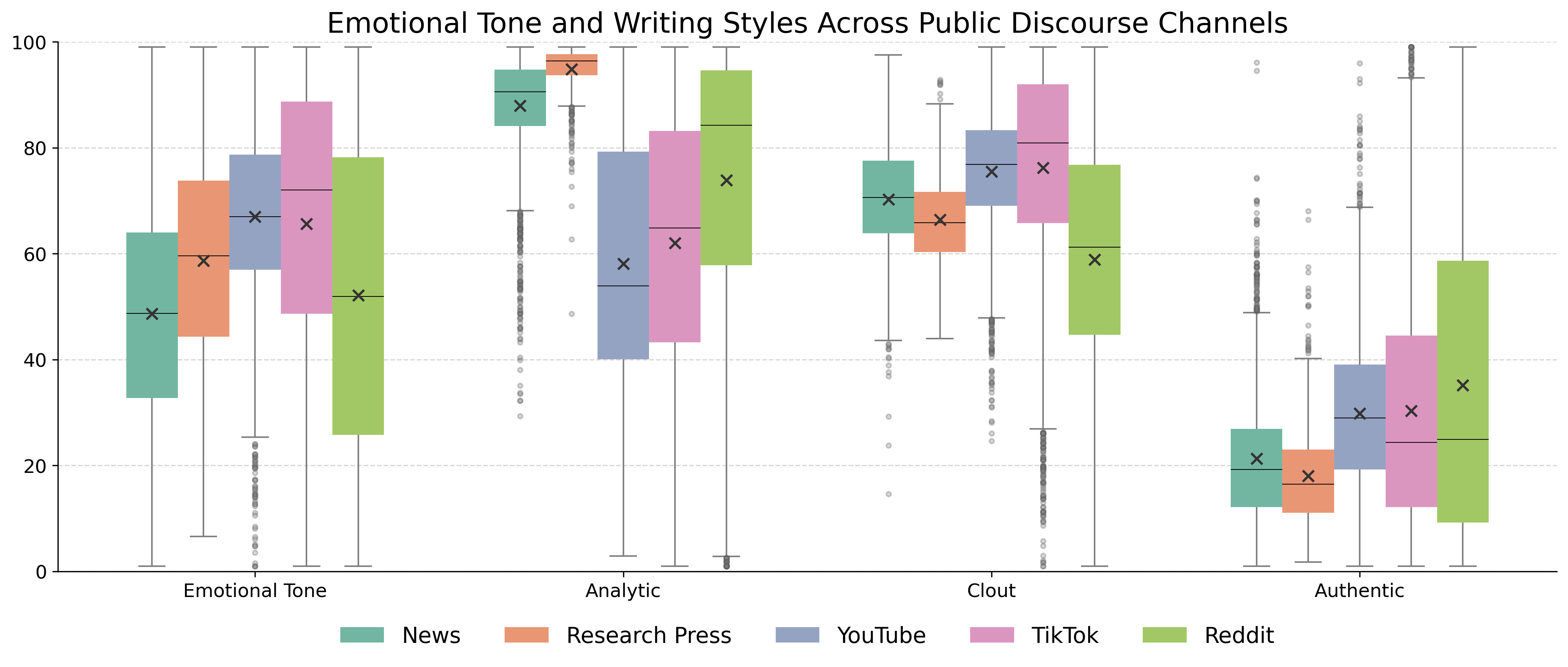}
\caption{Lexical style in public discourse on LLMs for health. Kruskal–Wallis H-tests were performed to determine whether there was a significant difference across channels (*** $p$<0.001, ** $p$<0.01, * $p$<0.05), where emotional tone $H$ = 1332.73 (***), analytic $H$ = 3552.45 (***), clout $H$ = 2895.86 (***), and authentic $H$ = 419.56 (***). Dunn's post hoc test indicated that content on YouTube and TikTok was significantly more positive, while news was significantly more negative than other channels.}
\label{fig:style}
\Description{A grouped boxplot compares News, Research Press, YouTube, TikTok, and Reddit across four measures: Emotional Tone, Analytic, Clout, and Authentic. Discourse is generally neutral to positive, with TikTok and YouTube showing a higher emotional tone and News and Research Press being more neutral. Analytic language is highest in Research Press and News, and more variable and lower for social media platforms. Clout is highest on TikTok and YouTube. Authenticity is highest on Reddit and lowest in News and Research Press.}
\end{figure*}

We observed differences in writing style across discourse channels (Fig.~\ref{fig:style}), which reflects the underlying communication goals and norms of each venue. 
First, News and Research Press displayed the highest level of analytical writing that is related to logical reasoning and formal communication. This tendency aligns with the expectation of rationality and professionalism in institutional communication. In comparison, the other three user-driven social media platforms exhibited lower and more varied levels of analytical style, which suggests a tendency for these social media platforms to prioritize accessibility over formal reasoning. 

The second dimension, clout, captures self-focused expressions that are associated with confidence and leadership. Contrary to analytical expressions, YouTube and TikTok had higher scores, indicating that content creators on these two platforms tended to present themselves as opinion-influencing figures. These two platforms tended to use more collective pronouns (e.g., \textit{``\textbf{we} love that ChatGPT thinks like \textbf{us}''}) or second-person singular pronouns (e.g., \textit{``so \textbf{you} have downloaded ChatGPT and played around with it. \textbf{you} liked it but \textbf{you}'re probably asking how do I save money with it''}), potentially because these platforms encourage videos with one-to-one conversational styles and expressions of shared feelings.
Dunn's post hoc test further demonstrated that YouTube and TikTok had significantly higher clout scores than other channels. Both platforms also used more certainty words (e.g., \textit{never, definitely}) and fewer tentative words (e.g., \textit{maybe, perhaps}); for instance, \textit{``they are the doctors you can \textbf{always} trust, \textbf{always} just a message away.''} 

Lastly, all public channels incorporated a lower level of authentic communication styles, suggesting they had a high degree to which a person is self-monitoring. This lower authentic tendency suggests that cognitive process words could be used to support statements, and the overall language is more abstract. Among them, News and Research Press had the lowest average levels, which can be explained by expectations for professional writing to avoid personal or informal language.

\subsection{Informational Content}
\label{result:content}

\begin{shaded}
\noindent\textbf{Summary:} (1) Public discussions of LLMs in health emphasized specific and isolated cases, with episodic framing more common than thematic framing. Practical implications for healthcare, safety, and the economy were highlighted more, while broader systemic or policy-related considerations were rarely touched on, except in the news. 
(2) Overall, public discourse lacked a thorough mention of risks, and the generative nature and distinctive affordances of LLMs were rarely discussed. When risks were discussed, information quality issues were most frequently highlighted. 
(3) Layperson-driven platforms were especially less likely to mention risks or explain generative nature than institutional sources, while YouTube served as a middle ground with both organization- and user-generated content.
(4) In terms of specific health subdomains, TikTok and Reddit emphasized consumer-facing applications, particularly mental health support, while News and Research Press highlighted clinical decision support and research-oriented applications. 
\end{shaded}

\subsubsection{Framing Type and Dimensions.}

\begin{figure}
\small
\centering
\begin{subfigure}{\columnwidth}
    \includegraphics[width=.85\columnwidth]{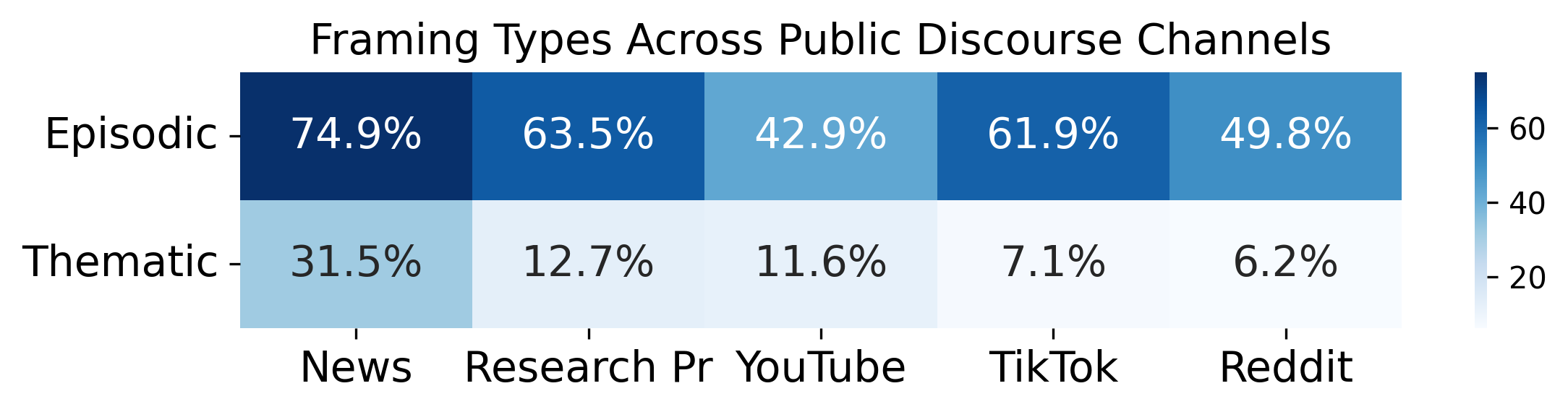}
    \centering
    \caption{Framing types (***): whether narratives are framed episodically or thematically.}
    \label{fig:framing_type}
    \Description{A heatmap shows the frequency of episodic and thematic framing across News, Research Press, YouTube, TikTok, and Reddit. Episodic framing dominated across all platforms: News 74.9\%, Research Press 63.5\%, TikTok 61.9\%, Reddit 49.8\%, and YouTube 42.9\%. Thematic framing was much less common overall, with News again highest at 31.5\%, while Research Press 12.7\%, YouTube 11.6\%, TikTok 7.1\%, and Reddit 6.2\% showed lower levels.}
\end{subfigure}
\begin{subfigure}{\columnwidth}
    \centering
    \includegraphics[width=\columnwidth]{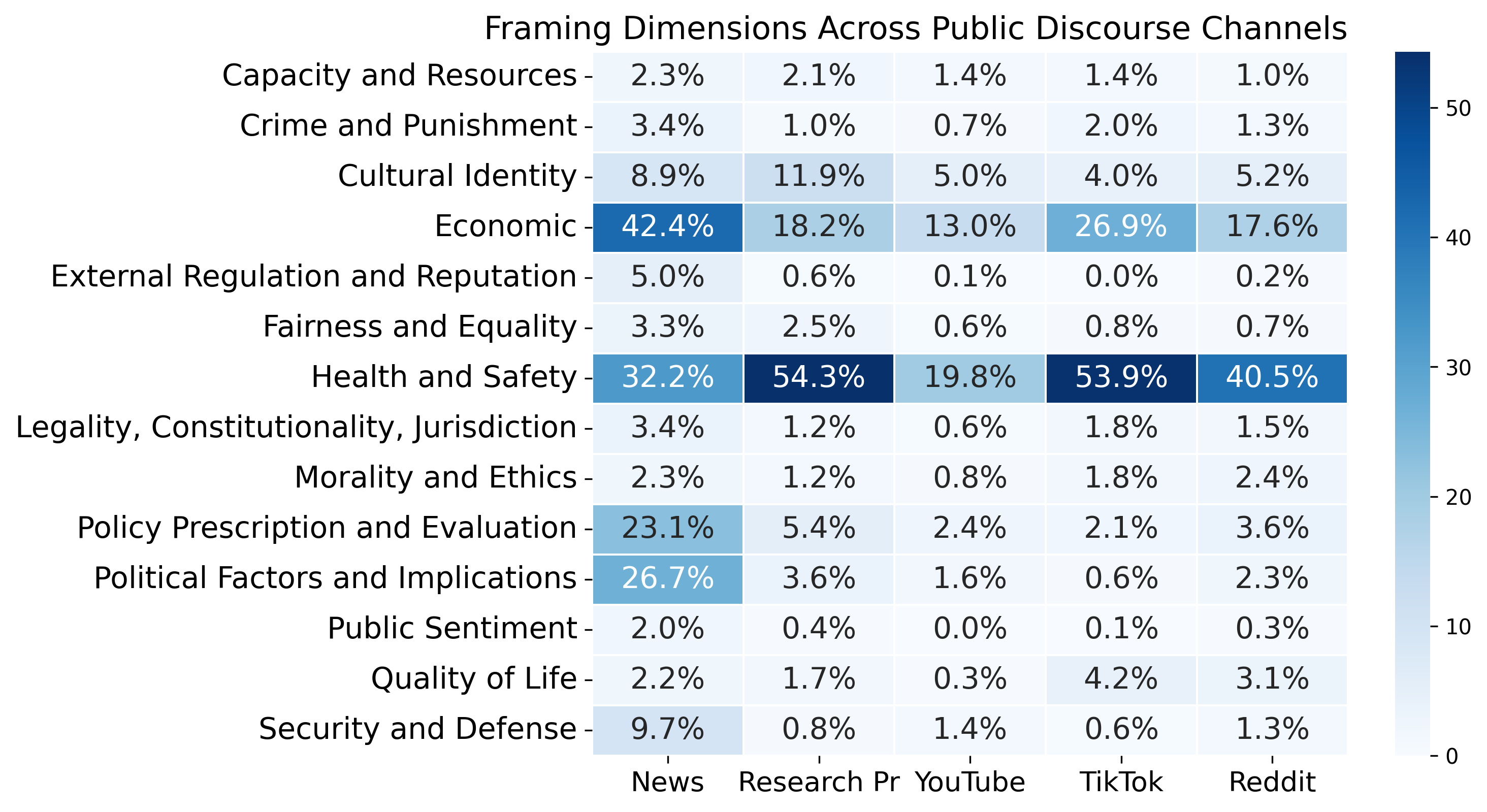}
    \caption{Framing dimensions (***): what broader policy-related themes are framed as relevant.}
    \label{fig:framing_dimension}
    \Description{A heatmap compares issue framing across News, Research Press, YouTube, TikTok, and Reddit. Health and Safety was the most discussed issue, especially on Research Press and TikTok, while Economic-related issues were most prominent in News and also notable on TikTok and Reddit. News uniquely features higher levels of Policy- and Political-related discussions. Most other frames (e.g., Crime, Fairness, Legality, Morality, Security) appear at low levels across all channels.}
\end{subfigure}
\caption{Framing types and dimensions in public discourse on LLMs for health. Kruskal–Wallis H-tests showed significant differences at all levels across channels (*** $p$<0.001, ** $p$<0.01, * $p$<0.05), with higher $H$ in thematic framing and discussions related to policy prescription and evaluation, as well as political factors and implications.}
\label{fig:framing}
\end{figure}

For \textit{framing type}, we found that episodic framing was noticeably more prevalent than thematic framing across all discourse channels (Fig.~\ref{fig:framing_type}). This means that LLM adoption in health is often introduced through specific and isolated cases, such as spotlighting particular use cases or technological milestones, rather than being presented as a generalized and systematic happening. Kruskal–Wallis H-tests showed significant differences in framing types across channels, where episodic $H$ = 715.82 (***) and thematic $H$ = 1310.74 (***). Notably, News showed a significantly higher proportion of thematic framing (31.5\%) than other sources, which may be explained by journalistic writing patterns that tend to situate events within larger societal contexts.

In terms of \textit{framing dimensions}, direct and practical impacts on healthcare, safety, and economy were more common, while broader systemic, ethical, or policy-related framings remain underrepresented, particularly in social media platforms (Fig.~\ref{fig:framing_dimension}). Unsurprisingly, given this study's focus on LLMs in health, health and safety had a higher prevalence, ranging from 19.8\% in YouTube to 54.3\% in Research Press. The next most frequently mentioned social issue dimension was economic, especially for News (42.4\%) and TikTok (26.9\%). A closer examination of data showed some subtle differences: News tended to emphasize corporate and macroeconomic implications, while TikTok had more discussions on personal financial impacts. Nevertheless, disruption of the labor market appeared as a consistent theme, ranging from personal worries (e.g., \textit{``I think I’m officially out of a job. I asked ChatGPT to make a knee rehab exercise plan, and it came up with this in seconds...''}) to broader societal impacts (\textit{``Although conversations surrounding technological unemployment over the past several decades have revolved around blue-collar workers losing their positions to automated robotics solutions, the widespread use of ChatGPT has introduced similar questions in the healthcare and other knowledge-based professions.''}).

Kruskal–Wallis H-tests showed significant differences in framing dimensions across channels, where capacity and resources $H$ = 28.44 (***), crime and punishment $H$ = 74.14 (***),  cultural identity $H$ = 102.03 (***), economic $H$ = 863.32 (***), external regulation and reputation $H$ = 680.83 (***), fairness and equality $H$ = 139.71 (***), health and safety $H$ = 962.78 (***), legality, constitutionality, jurisdiction $H$ = 68.51 (***), morality and ethics $H$ = 40.89 (***), policy prescription and evaluation $H$ = 1534.82 (***), political factors and implications $H$ = 2776.45 (***), public sentiment $H$ = 152.55 (***), quality of life $H$ = 117.41 (***), security and defense $H$ = 691.13 (***).

Overall, News was the only venue that tended to contextualize LLMs in health across a wide range of social issues, including economic, policy, political, safety, and cultural considerations. On the other hand, other platforms had less comprehensive discussions, with Reddit and TikTok showing the lowest coverage of dimensions like resource, policy, or fairness. Instead, they had a significantly higher tendency to mention the impacts on quality of life than other venues, suggesting their personal and practical perspective.


\subsubsection{Risk and Generative Nature Disclosure.}

\begin{figure}
\small
\centering
\includegraphics[width=\columnwidth]{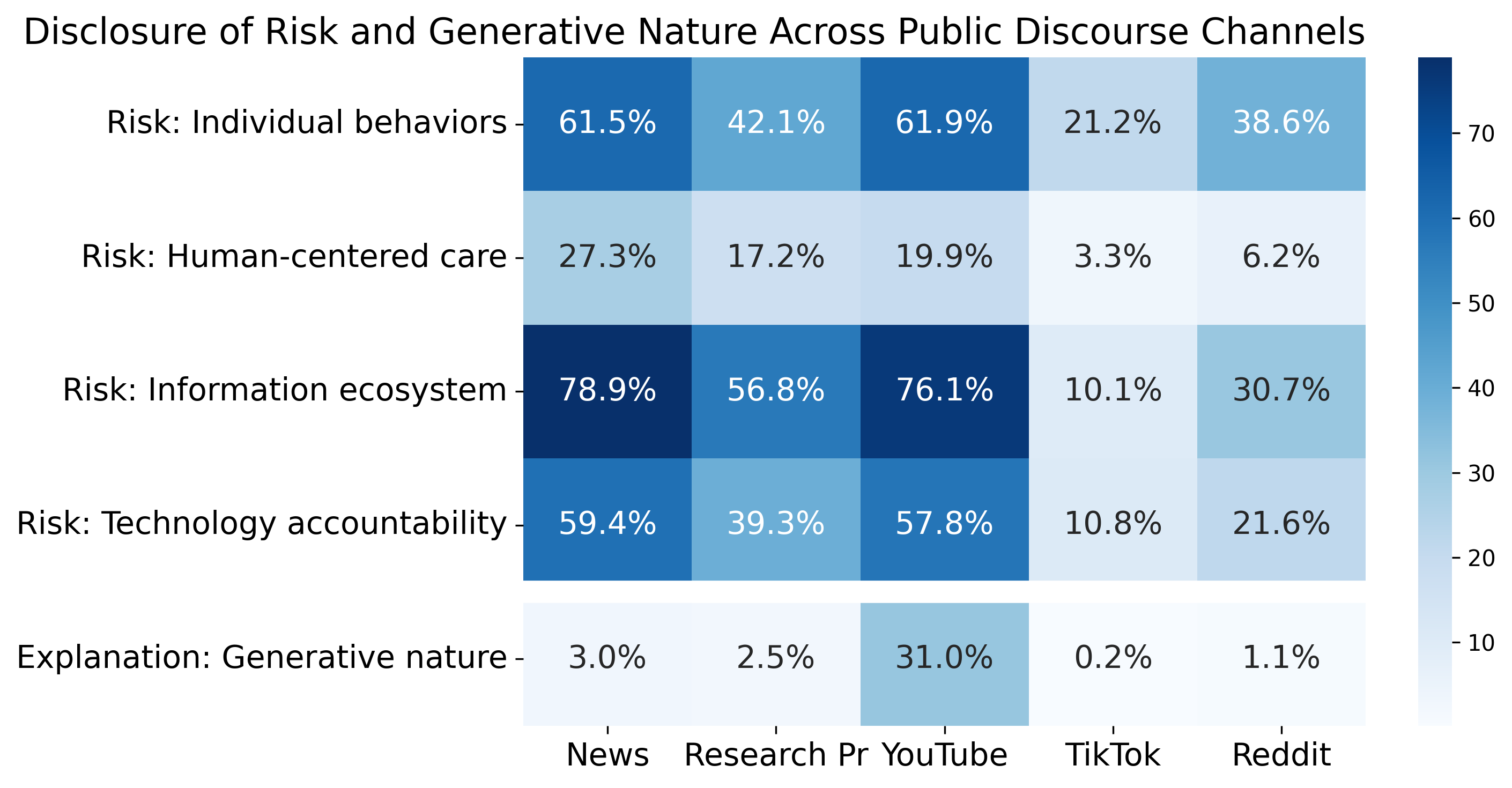}
\caption{Disclosure of risks and generative nature in public discourse on LLMs for health. Public discourse exhibited unbalanced risk disclosure across channels, lacked thorough risk coverage, and rarely discussed the generative nature and distinctive functionalities or affordances of LLMs. Kruskal–Wallis H-tests showed significant differences across channels (*** $p$<0.001, ** $p$<0.01, * $p$<0.05), where risks to individuals $H$ = 1564.17 (***),  risks to human-centered care $H$ = 1460.14 (***), risks to information ecosystems $H$ = 5042.14 (***), risks to technology accountability $H$ = 3213.67 (***), and explanations of generative nature  $H$ = 4618.17 (***).}
\label{fig:risk}
\Description{A heatmap shows how often different AI risks and explanations appear across News, Research Press, YouTube, TikTok, and Reddit. News and YouTube most frequently discuss information ecosystem and individual behavior risks, along with technology accountability. Research Press shows moderate attention to these risks, while TikTok and Reddit mention them less often. Explanations of AI’s generative nature are rare overall but appear much more on YouTube than on other platforms.}
\end{figure}

As shown in Fig.~\ref{fig:risk}, we found that public communication surrounding LLMs overall lacked a thorough introduction or overview to their risks, especially in explaining different mechanisms from traditional information sources. There was a notable asymmetry in the engagement of risk communication across venues, where layperson-driven platforms (i.e., TikTok and Reddit) were less likely to mention risks and LLM mechanisms, compared to professional and institutional sources such as News and Research Press.

The most frequently addressed risk across all platforms was related to the \textit{information ecosystem} --- how LLMs may affect people, technologies, and norms that shape the way information is produced, shared, and evaluated. A closer examination of the data showed that this type of concerns was concentrated on direct and isolated information quality issues. Common examples include hallucination and inaccuracies (e.g., plausible-but-wrong answers or fabricated citations), bias (e.g., racist outputs), or misuse of AI-generated content (e.g., deepfakes or AI-generated misinformation). For example, \textit{``I asked ChatGPT how to get rid of chronic pain from meniscus injuries. While its suggestions weren't completely wrong, they weren't exactly right either, and can worsen the situation.''} This type of information quality issue was especially commonly mentioned in News (78.9\%), YouTube (76.1\%), and Research Press (56.8\%).

The next two most commonly mentioned categories were risks to individual behaviors and risks to technology accountability, and risks to human-centered care were the least frequently mentioned.
\textit{Risks to individual behaviors} included potential negative influences on individuals' actions, decision-making, and well-being. This type of risk was frequently mentioned in News (61.5\%) and YouTube (61.9\%). Common examples included impacts on employment and work norms, over-reliance on AI for decision-making, and the erosion of interpersonal or professional skills.

Similarly, \textit{risks to technology accountability} were less prevalent on TikTok and Reddit, despite appearing with reasonable frequency in other venues. These risks refer to negative consequences related to how LLMs may challenge existing mechanisms of oversight and security. Some frequently discussed examples were algorithmic biases, unauthorized imitation of personal traits or proprietary work, and misuse for harmful purposes or human-like roles.
Last, \textit{risks to human-centered care} were least frequent. These concerns surround how LLMs may undermine the relational foundations of healthcare and disrupt existing social support systems. Despite all content touching on LLM's potential in health, risks to the healthcare system were infrequently addressed in institutional sources and nearly absent on TikTok (3.3\%) and Reddit (6.2\%).

One concerning observation in our results is that explicit explanations of LLMs' generative nature, which makes LLMs differ from conventional health information technologies, were nearly absent in all channels. Only YouTube made a moderate mention of mechanisms (31.0\%), likely due to its longer-form and tutorial-based content. Across News, Research Press, Reddit, and TikTok, references to the workings of LLMs remained less than 3\% or near absent (Reddit 1.1\% and TikTok 0.2\%), indicating a widespread lack of conceptual differentiation between LLMs and traditional search engines or decision aids -- popular tools for online health information seeking in the past two decades.

\subsubsection{Health Subdomain Prevalence}

\begin{figure*}
\small
\centering
\includegraphics[width=0.75\textwidth]{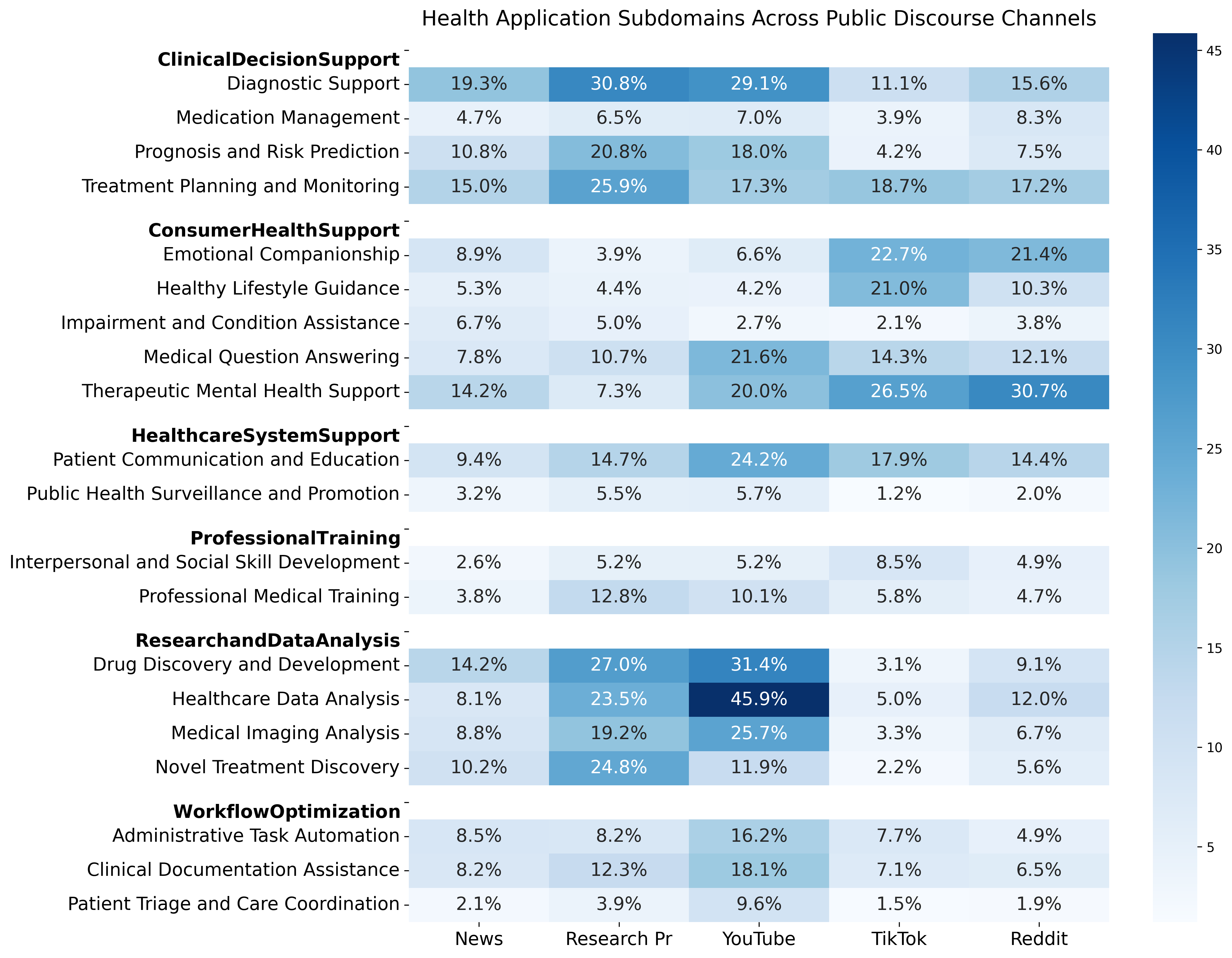}
\caption{Health subdomains in which LLMs are mentioned in public discourse. TikTok and Reddit emphasized consumer-facing applications, particularly mental health support, while News and Research Press focused on clinical decision support and research-oriented uses. YouTube's longer, education-oriented content contributes to broader topical coverage. Kruskal–Wallis H-tests showed significant differences only in Research and Data Analysis, where $H$ = 15.76 (**) (*** $p$<0.001, ** $p$<0.01, * $p$<0.05).}
\label{fig:subdomain}
\Description{A heatmap compares health-related AI application subdomains across News, Research Press, YouTube, TikTok, and Reddit. Clinical decision support and research and data analysis were most common in Research Press and YouTube. Social platforms, especially TikTok and Reddit, emphasized consumer health uses such as emotional companionship and mental health support. Workflow and system-level applications appeared at moderate levels, while professional training and public health topics were relatively rare across all channels.}
\end{figure*}

We identified what health subdomains in which LLMs are presented as applicable in public discussion from the inductive coding. We summarized 20 subcategories that were organized into six broader categories: clinical decision support, workflow optimization, professional training, research and data analysis, consumer health support, and healthcare system support. As shown in Fig.~\ref{fig:subdomain}, the most commonly mentioned health subdomains were concentrated in clinical decision support (especially in diagnosis support and treatment planning and monitoring) and consumer health support (particularly in therapeutic mental health support and medical question answering). Meanwhile, the least discussed subdomains were those related to professional training and workflow optimization.

Among platforms, YouTube emerged as a middle ground between institutional sources (e.g., News and Research Press) and layperson-driven social media platforms (e.g., TikTok and Reddit). It tended to play a unique role with a large number of educational and business materials, such as talks, lectures and tutorials, which was reflected in its significantly higher content length (see Table~\ref{table:data}) and could contribute to its broader topical coverage. This explains its focus on practical use cases, especially in healthcare data analysis, drug discovery, diagnostic and risk prediction, and medical imaging analysis. YouTube's relatively even distribution across clinical, research, and support-oriented subdomains indicates its unique positioning that blends professional and public-facing communication.

On the other hand, the other two social media platforms, TikTok and Reddit, showed clear emphasis on consumer-facing and mental health support applications. The most frequently mentioned areas were therapeutic mental health support (TikTok: 26.5\% and Reddit 30.7\%) and emotional companionship (TikTok: 22.7\% and Reddit 21.4\%). Followed by treatment planning and monitoring, healthy lifestyle guidance, and patient communication and education. Aligned with their tendency to focus on personal and practical perspectives as observed in the framing patterns, they were less likely to discuss LLM applications in research and data analysis.

News, by comparison, emphasized clinical decision support, especially diagnostic support and risk prediction, alongside notable attention to therapeutic mental health support. Research Press followed a similar trend in highlighting clinical applications, but also had substantial coverage of research-oriented subdomains, such as drug discovery, healthcare data analysis, and novel treatment discovery, aligned with its unique role in communicating scientific and translational research findings.

\subsection{Symbolic Representation}

\begin{shaded}
\noindent\textbf{Summary:} Public discussions had relatively neutral to low overall anthropomorphism of LLM entities. TikTok and Reddit showed wider ranges of anthropomorphism, with some highly human-like portrayals of LLMs, while Research Press showed the significantly least anthropomorphism.
\end{shaded}

\begin{figure}
\small
\centering
\includegraphics[width=\columnwidth]{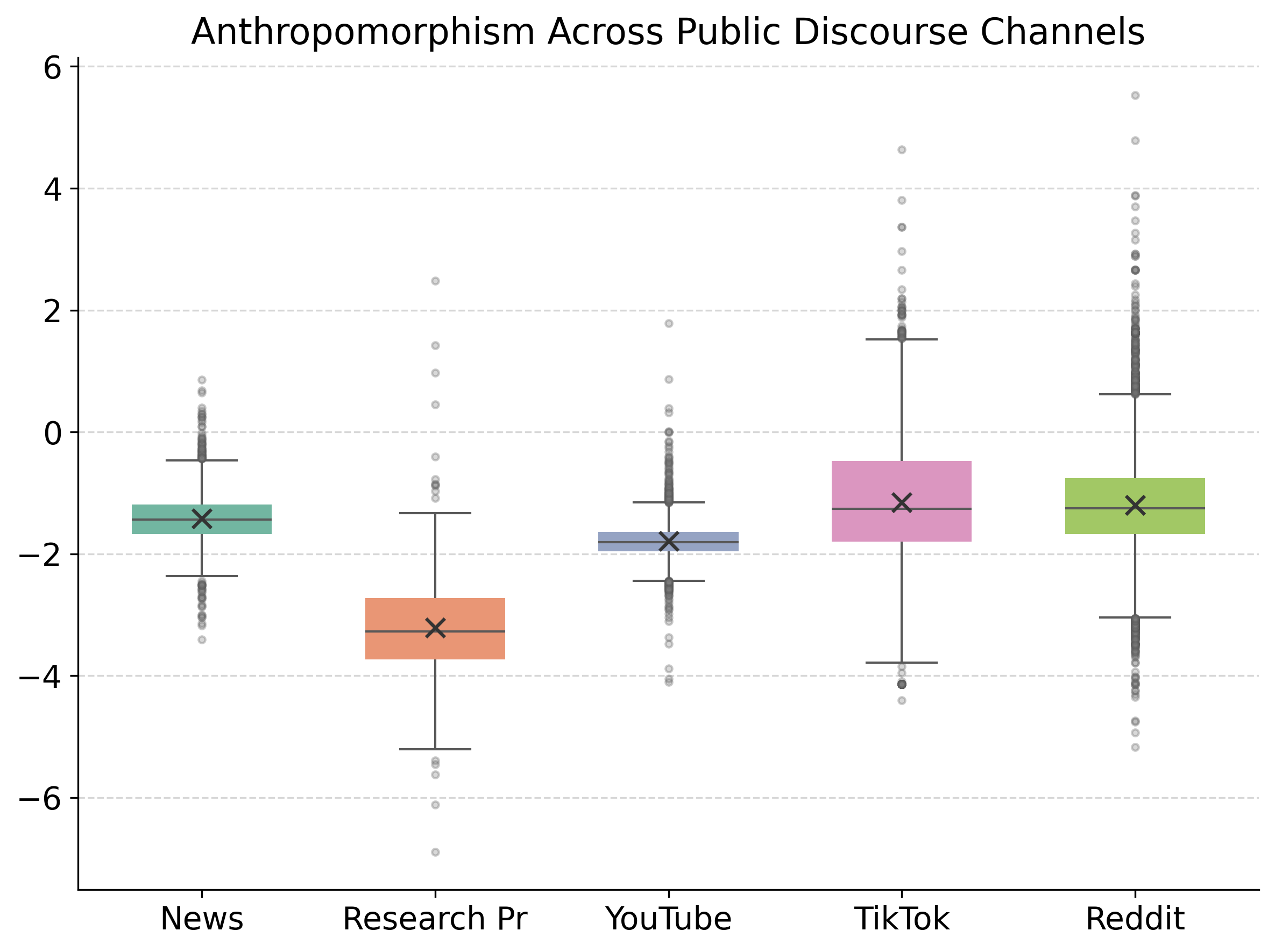}
\caption{Anthropomorphism of LLM entities in public discourse on LLMs for health. Public discussions had relatively neutral to low overall anthropomorphism of LLM entities, with some highly human-like portrayals of LLMs in TikTok and Reddit. Kruskal–Wallis H-test found significant differences across channels, where $H$ = 3974.28 (***). Dunn's post hoc test indicated that News was significantly less likely to adopt anthropomorphized representations of LLM entities.}
\label{fig:anthro}
\Description{Boxplots compare anthropomorphic language across News, Research Press, YouTube, TikTok, and Reddit. Anthropomorphic scores were generally below zero on all platforms, indicating limited anthropomorphism overall. Research Press showed the lowest levels, News and YouTube were slightly higher, and TikTok and Reddit showed comparatively more anthropomorphic language and greater variability.}
\end{figure}

In examining the symbolic representation of LLMs, we measured the degree of anthropomorphism in describing LLM-related entities. As Fig.~\ref{fig:anthro} shows, expressions across all discourse channels had interquartile ranges below zero, which means that public discussions had no apparent anthropomorphism tendencies and most direct references framed LLMs more as non-human entities.
At the same time, we observed cross-platform differences in the representation of AI. Specifically, Research Press exhibited the lowest levels of anthropomorphism, which aligns with its tendency to adopt technical descriptions. In contrast, TikTok and Reddit tended to have wider ranges of anthropomorphism scores. These two platforms also contained highly anthropomorphized expressions with scores near or above 2 that described LLMs with human-like capabilities (e.g., \textit{``my ChatGPT therapist just knows what's up in my mind''}), intentions (e.g., \textit{``my chatbot cares about me''}), or relationships (e.g., \textit{``he fell in love with and married his AI bot''}). 
Overall, verbs associated with the highest anthropomorphism scores included \textit{``arrive'', ``terrify'', ``baffle'', ``learn''}, while common verbs mentioned in the lowest scores were \textit{``cost'', ``apply'', ``use'', ``include''}. On the other hand, News and YouTube showed an intermediate pattern of moderate and less variable anthropomorphic framing, with fewer extreme values.


\section{Discussion}

This paper investigates how public discourse introduces large language models, and frames their potential uses and risks in the health domain: one of the most discussed and consequential application areas. Overall, we found that discussions on LLMs for health across platforms were generally positive --- and becoming more so over time, but often lacked a thorough introduction or overview of risks. When risks were mentioned, they were largely about information quality, with nearly absent explanations about the generative nature of LLMs that distinguish them from traditional information sources. 
Across channels, layperson-driven platforms such as TikTok and Reddit were less likely to mention risks and displayed greater variations in emotional tone and anthropomorphic expressions. They also emphasized consumer health (especially mental health support), whereas professionally authored sources, such as news and research press, put greater emphasis on clinical use and were more likely to mention broader societal or systemic implications.

Together, this work shows that discourse can work as a diagnostic tool to track public expectations and attention, as well as to identify gaps in literacy or governance. In the case of LLM-related discussions, public attitudes were found to be generally positive and increasingly so over time, but risk coverage was limited and explanations of LLMs' generative nature were nearly absent. These findings underscore a need for effective communication and design strategies to help the public build agency and knowledge in navigating engagements with LLMs and their applications in health.

\subsection{The Role of Public Discourse in Negotiating AI Perceptions}

One of our primary contributions is a large-scale empirical analysis of how LLMs for health are represented and framed across diverse media channels, which we believe serves as an infrastructure for shaping public perceptions and negotiating practices. Prior research has demonstrated the role of media in communicating the risks and benefits of emerging technologies~\cite{boholm1998comparative, marks2007mass, weaver2009searching}. In terms of AI risks, scholarship has found that the media coverage, ranging from newspapers and magazines to blogs, had a shallow and mainly practical focus on AI ethics issues~\cite{ouchchy2020ai}, while global news showed a skewed prioritization on societal and legal and rights-related risks~\cite{allaham2025informing}.
Our findings extend these by documenting how different channels of public narratives reflect and reinforce this effect. Across platforms and content creators, we found that discussions were largely positive while notably lacking in mentioning risks or explaining the generative nature of LLMs. This is concerning as LLMs' functionalities (e.g., probabilistic generator) and affordances (e.g., general-purpose and non-task-driven design) mechanistically differ from traditional information sources. 

The lack of discussion of risks and explanation of the distinctive functionalities of LLMs can encourage individuals to interact with LLMs using ``old norms'' that may no longer apply. Particularly at a time when current design patterns of many LLM systems often overshadow the generative nature of AI outputs or blur the boundaries between generative and retrieval systems.
As technological affordances will inevitably shape users' mental models~\cite{Norman_design}, recent empirical findings showed that people held inaccurate perceptions of LLMs~\cite{zhang2024s, zhou2024risk}. Taken together, all evidence underscores the importance of efforts to regulate public perception and promote AI literacy. While recent regulations like the European Union (EU) AI Act~\cite{eu_ai_act_2024} have emphasized the importance of AI literacy, these efforts focused on programs ``intended for professionals, organisations, or the general workforce''~\cite{eu_ai_act_literacy}. Thus, we highlight the unique role of public discourse in engaging and influencing the greater public's everyday perceptions and practices. After all, communicating risks should be a two-way exchange of information~\cite{keeney1986improving}, not reduced to correcting ``wrong'' concerns or edifying ``uninformed'' lay perceptions. 

A closer look at cross-channel differences reveals another important feature of the discourse infrastructure: heterogeneity among content author types. Our results showed that professional outlets (i.e., news and research press) were more likely to mention risks and broader societal and systemic implications beyond specific and isolated instances. Meanwhile, there was also an unaligned focus on specific health domains for LLMs applications in layperson outlets: TikTok and Reddit had more discussions surrounding consumer health support and especially in emotional companionship and therapeutic support. 
This asymmetry has implications for both information environments and information consumers. For \textit{information environments}, it suggests that areas of high public interest (here mental health care) were under-addressed by professional outlets, leaving other channels and content that are more risk-agnostic to fill the gap. This highlights the need for journalists, researchers, and professionals to actively identify where public interest is situated, and to develop strategies for reaching audiences in those spaces with more balanced communication strategies. For \textit{information consumers}, relying primarily on platforms like TikTok and Reddit may reduce exposure to risk-aware perspectives and increase the likelihood of developing over-optimistic or anthropomorphized views of AI. Here, a critical awareness of platform differences and tendencies becomes essential, so as to allow audiences to interpret content not only based on what is said, but also in light of the communication environments that shape what is highlighted, appreciated, or encouraged.

On a higher level, the greater variation in tone and anthropomorphism in TikTok and Reddit suggests that public perceptions of LLMs remain unsettled and actively negotiated. We see this active negotiation as both a risk and an opportunity. On one hand, selective exposure to highly positive or human-like portrayals can skew expectations and normalize inappropriate mental models. On the other hand, these unsettled framings can provide insight for journalists, designers, and policymakers to understand where public attention is gathering, where professional communication is absent, and where new gaps in literacy or governance emerge.

\subsection{Implications for AI Transparency and Governance}

\subsubsection{What to disclose in supporting AI transparency.}
In efforts to support AI transparency and literacy, the first question that follows is ``what to disclose or educate''. Current regulations offer some direction: for instance, the EU AI Act requires that ``developers and deployers must ensure that end-users are aware that they are interacting with AI (chatbots and deepfakes)''~\cite{eu_ai_act_2024}. While such disclosure is important, especially as AI systems are increasingly embedded into existing systems and ``disappear into'' prior technologies, it may not be sufficient. Especially when categories of AI have not reached consensus in public understanding, end-users lack clarity about what capabilities and limitations different systems entail or offer. Our findings raise the question of whether informing people that they are ``interacting with AI'' meaningfully prepares them to interpret and evaluate their interactions.

Prior literature on risk communication underscores that effective messaging should go beyond stating the existence of risk; instead, it involves communicating not only the nature of risks but also expressing concerns, perspectives/opinions, and reactions to them~\cite{keeney1986improving}. Some current common practices include acknowledging error possibility~\cite{hill2025_psychosis} and uncertainty indicators~\cite {kim2024m} or reminding limitations of language models~\cite{wester2024ai}. But past work in human-AI interaction has emphasized both functional understanding of what models or systems work and mechanistic understanding of how models or systems work in AI transparency~\cite{liao2023ai}, and showed that analytical thinking that disrupts heuristic thinking on direct statements can reduce AI overreliance~\cite{buccinca2021trust}. 

However, turning to our findings, we will find a significant gap: explanations about the generative nature of LLMs were nearly missing across all channels. Even when risks were mentioned, they overwhelmingly centered on information quality, and a closer look at the data shows that the majority of the content focused on isolated incidents of fake or incorrect information. However, we question whether collapsing all information into the frequently mentioned yet broad category of hallucination is enough, when the metaphor ``hallucination'' itself is imperfect. LLMs do not have sensory perception as humans, so it does not naturally carry the appropriate referential analogy to explain why such errors occur. Therefore, we call attention to the gap in actively disclosing and educating on the mechanistic understanding of LLMs: how they differ from prior technologies and why these differences may lead to risks.

\subsubsection{What to align in facing governance lag.}
In examining the specific health subdomains mentioned in pubic discussions, we found that these health application subdomains align broadly with categories in World Health Organization's guidance on large multimodal models~\cite{world2024releases}, but offered more nuanced typology in specific use cases. Importantly, as mentioned above, we found that compared to professional-authored content (news articles and research press releases) that emphasized clinical or systemic uses, TikTok and Reddit mentioned consumer health and mental health care more --- ranging from emotional companionship to therapeutic support. This in part could be attributed to the ongoing mental health crisis in the United States and globally, especially among youth populations~\cite{cdc2024youth, who_adolescent_mental_health}, and their pursuit of support. For instance, a survey in 2025 found that 72 percent of American teenagers said they had used AI as companions~\cite{csm2025_teen}. Alongside this trend, researchers have documented alarming risks, as LLMs can encourage schizophrenic delusions~\cite{Landymore2024psychologist}, self-harm and abusive behaviors~\cite{CCDH2025_fake, wilkins2024teen}, and inability to handle emergency or high-stakes situations~\cite{wang2021evaluation, kim2024mindfuldiary}, and even being linked to actual suicides or violent crimes~\cite{dupre2025kill, wilkins2024teen}.

These documented risks, combined with our findings on high demands, point to the blurred line between informal emotional companionship and high-stakes therapeutic interventions in AI systems, which can raise safety concerns that policy frameworks have only begun to address. Some recent governance efforts in specifying boundaries of AI use for mental health care include Illinois' Act that prohibits AI from acting as a therapist or counselor~\cite{gil2025illinois} and New York's requirement for AI tools to refer people with suicidal or self-harm tendencies to crisis centers~\cite{NY2025_ai}. However, overall, research, practice, and policy in this area are emergent and rapidly evolving, as professional societies and stakeholders are urging AI tools to be grounded in psychological science, co-developed with behavioral health experts, and rigorously tested for safety~\cite{APA2025_chatbots}.

This misalignment between public demand and regulatory clarity underscores the need for future work on both design and governance. Future work can develop strategies for in situ literacy through scaffolding or heuristic interactions to prepare users in developing mechanistic understanding, evaluating outputs, and critically trading off risks and benefits in sensitive contexts like health. On the governance side, empirical evidence from public discourse can help regulators prioritize domains where demand is high and risks are pressing or imminent. Literature in risk communication, mass media, and health education all suggest that the more effective approach is to promote better understanding and greater agency and stewardship~\cite{das2018breaking, keeney1986improving, janz1984health}. We envision opportunities for future work to contribute experimental and empirical evaluations to provide better clarity as to what information most effectively supports user agency and informed engagement.


\subsection{Limitations and Future Work}
From a generalization perspective, this work focused on English-language content with some data (TikTok, YouTube, and News) limited to the U.S context. While some patterns, such as the positive framing of LLMs or limited discussion of risks, can be observed in other cultural or linguistic contexts (e.g., ~\cite{allaham2025informing}), we acknowledge that not all findings can be directly applied to non-English or non-U.S. discourse. Future work can extend this analysis to multilingual and cross-cultural settings to better understand global public perceptions of LLMs. Similarly, our categorization of risks was based on one taxonomy tailored to LLMs in the health domain. While this choice aligns with our study focus, it may not capture all potential risk dimensions relevant to other domains, and future work can utilize the methodological framework of this study to analyze public discourse on LLMs in other domains. 
Second, our data and analyses may be subject to platform- and source-related biases: TikTok API cutoffs may skew results toward older content, Pushshift may incompletely capture Reddit activity and introduce sampling bias, and MBFC filtering may favor mainstream outlets. In addition, although cross-model validation demonstrated the reliable performance of \texttt{gpt-4o-mini}, our large-scale annotations rely on this single model and may still contain model-specific biases. Third, even with human validation, we acknowledge that human annotations of complex constructs such as health subdomains and framing remain subject to interpretive uncertainty~\cite{card2015media} and automated labeling is inherently imperfect. Therefore, these labels should be interpreted as approximate signals rather than ground truth.
Lastly, we drew on the agenda-setting theory and focused on the salience of issues and certain attributes in messages. There are additional subtle attributes that can affect audience interpretation and attitude, such as visual elements in videos and varying levels of emphasis or depth in the presentation of information elements. Mentions of LLM capabilities, limitations, or risks were treated equally in this work, regardless of whether they appeared early or late in a narrative, or were introduced superficially or in detail. We hope our study can inspire future work to examine more fine-grained aspects within discourse content and their effects on perceptions of technological capabilities and risks.


\section{Conclusion}
This paper examined how public discourse introduces large language models (LLMs) in the health domain, an area where concerns about reliability are emerging alongside reports linking LLM use to risky behaviors, psychological risks, and even suicide cases. Analyzing five channels (news, research press, YouTube, TikTok, and Reddit) over a 2-year period, we found that public discourse was generally positive and episodic, without risks being thoroughly communicated and often reduced to information quality incidents, and explanations of LLMs' generative nature were almost absent. Professional outlets more often addressed clinical and systemic implications, whereas layperson platforms of TikTok and Reddit highlighted mental health care and companionship, and showed greater variation in tone and anthropomorphism but little attention to risks. These findings position public discourse as a diagnostic tool for identifying where public attention gathers, where professional communication is absent, and where gaps in literacy or governance emerge. Our work contributes empirical evidence of how LLMs are currently framed in discourse and points toward future efforts to 1) support professionals in addressing gaps in information environments and information consumers in developing critical awareness of source differences; 2) develop communication strategies and in situ literacy scaffolds that move beyond surface disclosures to foster mechanistic understanding that help users critically assess capabilities and risks; and 3) guide regulatory efforts into areas with high demand but pressing risks.

\begin{acks}
Zhou and De Choudhury were partly supported through a contract from the U.S. Centers for Disease Control and Prevention (CDC). Findings reported in this paper represent the views of the authors, and not of their employers or the sponsor CDC. We thank Viet Cuong (Johnny) Nguyen, Shravika Mittal, Koustuv Saha, the Social Dynamics and Well-Being Lab members, and the anonymous reviewers for their valuable input on the paper.
\end{acks}


\bibliographystyle{ACM-Reference-Format}
\bibliography{reference}

\appendix

\section{Keywords Used in Public Discourse Data Collection}
\label{appendix:keywords}

\vspace{1em}
\textbf{LLM-relevant Keywords:} 

\noindent language model, llm, language artificial intelligence, language ai, generative ai, chatgpt, gpt, med-palm, google gemini, google bard, google lamda, transformer model, natural language generation, nlg, self-supervised learning, zero-shot, few-shot, generative model, autoregressive transformer, instruction tuned, meta llama, meta ai llama, meta galactica, microsoft copilot, mixtral, mistral ai, anthropic claude, deepseek, diffusion model, bloom model, qwen, xai grok

\vspace{2em}
\noindent \textbf{Health-relevant Keywords:}

\noindent \textbf{General and Symptom Terms:} health, medical, disease, well being, well-being, therapy, therapist, diagnosis, diagnose, drug, nutrition, clinic, clinical, medicine, doctor, surgeon, sick, illness, disorder, patient, mental, surgical, surgery, insurance, hospital, abuse, pharma, over dose, sore, tenderness, arm pain, back pain, breast pain, ear pain, ear ache, eye irritation, flank pain, general aches, general pain, body discomfort, hand pain, headache, head pain, head discomfort, kidney pain, knee pain, leg pain, neck-skull pain, rib pain, shoulder pain, sore throat, tooth ache, fatigue, sleep disturbance, abnormal dreams, vivid dreams, asthenia, disturbed sleep, insomnia, drowsiness, lassitude, listlessness, exhaustion, lethargic, sluggishness, somnolence, tiredness, weariness, abdominal pain, abdominal cramp, anorexia, changes in appetite, bloating, constipation, difficulty swallowing, dry mouth, dyspepsia, heartburn, indigestion, epigastric pain, upset stomach, nausea, stomach ache, stomach pain, affect lability, emotional instability, irritable mood, mood changes, mood swing, depression, depressed mood, suicidal ideation, blurred vision, neuropathy, pins and needles, tingling, dizziness, lightheadedness, head rush, wooziness, angina, chest pain, difficulty breathing, dyspnea, short of breath, congestion, heart flutter, palpitation, arthralgia, joint pain, muscle pain, stiffness, amnesia, brain fog, memory loss, difficulty concentrating, short attention span, skin irritation, itchiness, pruritus, skin discomfort, dysuria, vaginal dryness, sexual dysfunction, malaise, clinician, physician, nurse, surgeon, doctor, therapist, psychology, psychological, anxiety, depression, mental health, mental illness, mental disorder, bipolar, bpd, ptsd, paranoia, schizophrenia, schizophrenic, schizo, panic attack, anxiety attack, social anxiety, self harm, self-harm, eating disorder, binge eating disorder, anorexia, anorexic, bulimia, bulimic, depressed, depressing, suicidal, suicide
    
\noindent \textbf{Health Topics:} Abortion, Addictive behaviour, Adolescent health, Anaemia, Antimicrobial resistance, Assistive technology, Biologicals, Blood products, Blood transfusion safety, Brain health, Buruli ulcer (Mycobacterium ulcerans infection), Cancer, Cardiovascular diseases, Cervical cancer, Chagas disease, American trypanosomiasis, Chemical incidents, Chemical safety, Chikungunya, , Child health, Childhood cancer, Cholera, Chronic respiratory diseases, , Clinical trials, Common goods for health, Congenital disorders, Contraception, Coronavirus disease (COVID-19), Crimean-Congo haemorrhagic fever, Deafness and hearing loss, Dementia, Dengue and severe dengue, Depression, Diabetes, Diagnostics, Diarrhoea, Digital health, Diphtheria, Disability, Dracunculiasis (Guinea-worm disease), Injuries, Drugs, Earthquakes, Ebola virus disease, Echinococcosis, Epilepsy, Eye care, vision impairment and blindness, Female genital mutilation, Food fortification, Food safety, Foodborne diseases, Foodborne trematode infections, Health Laws, Behavioural interventions, Health promoting schools, Behavioural interventions, Health and well-being, Health promotion, Health technology assessment, Health workforce, Healthy diet, Hendra virus infection, Hepatitis, HIV, Hospitals, Human African trypanosomiasis (sleeping sickness), Human genome editing, Hypertension, In vitro diagnostics, Infant nutrition, Infection prevention and control, Infertility, Influenza (avian and  zoonotic), Influenza (seasonal), Infodemic, Intellectual property and trade, International Health Regulations, Landslides, Lassa fever, Lead poisoning, Leishmaniasis, Leprosy (Hansen disease), Lymphatic filariasis (Elephantiasis), Malaria, Malnutrition, Marburg virus disease, Maternal health, Measles, Medical devices, Medicines, Meningitis, Mental health, Micronutrients, Middle East respiratory syndrome coronavirus (MERS-CoV), Mpox, Mycetoma, chromoblastomycosis and  deep mycoses, Neglected tropical diseases, Newborn health, Nipah virus infection, Nursing and midwifery, Nutrition, Obesity, Occupational health, Onchocerciasis (river blindness), Oral health, Palliative care, Patient safety, Pertussis, Physical activity, Plague, Pneumonia, Poliomyelitis (polio), Primary health care, Quality of care, Rabies, Radiation, Radiation emergencies, Radon, Refugee and migrant health, Rehabilitation, Respiratory syncytial virus, Rift Valley fever, Road traffic injuries, Scabies, Schistosomiasis (Bilharzia), Self-care for health and well-being, Sepsis, Severe Acute Respiratory Syndrome (SARS), Health and wellbeing, Sexual and reproductive health and rights, Sexual health, Sexually transmitted infections (STIs), Smallpox, Snakebite envenoming, Social determinants of health, Soil-transmitted helminthiases, Stillbirth, Substandard and falsified medical products, Suicide prevention, Sustainable development, Syphilis, Taeniasis and cysticercosis, Tetanus, Tick-borne encephalitis, Tobacco, Trachoma, Traditional, Complementary and Integrative Medicine, Transplantation, Travel and health, Tropical Cyclones, Tsunamis, Tuberculosis, Typhoid, Ultraviolet radiation, Universal health coverage, Urban health, Vaccines and immunization, Violence against children, Violence against women, Volcanic eruptions, Women's health, Yaws (Endemic treponematoses), Yellow fever, Zika virus disease

\begin{table*}
\section{Human–Model Validation and Cross-Model Agreement of LLM Annotations}
\label{appendix:model_agreement}
\centering
\sffamily
\small
\begin{tabular}{lccccccccc}
     &  \multicolumn{3}{c}{\textbf{Risk Disclosure}}&  \multicolumn{3}{c}{\textbf{Generative Nature}}&  \multicolumn{3}{c}{\textbf{Health Subdomain}} \\
    \cmidrule(lr){2-4}\cmidrule(lr){5-7}\cmidrule(lr){8-10}
     \textbf{Model}&  F1&  Precision&  Recall&  F1&  Precision & Recall&  F1&  Precision&  Recall\\
     \hline
     gpt-4o-mini performance&  0.743&  0.725&  0.795&  0.851&  0.955& 0.800&  0.792&  0.745& 0.888\\
     gemini-2.5 performance&  0.715&  0.763&  0.698&  0.851&  0.955& 0.800&  0.742&  0.804& 0.705\\
     cross-model agreement& 0.710& 0.701& 0.744& 0.707& 0.783& 0.671&  0.775&  0.782& 0.768\\
\end{tabular}
\caption{Performance of GPT-4o-mini and Gemini-2.5 against human annotations (n = 100) and inter-model agreement on a larger subset (n = 500). All scores are macro-averaged.}
\end{table*}



\clearpage
\begin{table*}
\centering
\sffamily
\footnotesize
\section{Platform pairwise comparison}
\begin{tabular}{llrrclrrcl}
\toprule
\multicolumn{2}{l}{} & \multicolumn{4}{c}{\textbf{Tone}} & \multicolumn{4}{c}{\textbf{Analytic}}\\
&  &    & $d$ & 95\% CI & $p$ &    & $d$ & 95\% CI & $p$ \\
\hline
News & Research Pr & \scalebar{ -0.46 } & -0.46 & [-0.55, -0.37] & *** & \scalebar{ -0.75 } & -0.75 & [-0.84, -0.66] & *** \\
News & YouTube & \scalebar{ -0.96 } & -0.96 & [-1.02, -0.91] & *** & \scalebar{ 1.58 } & 1.58 & [1.52, 1.64] & *** \\
News & TikTok & \scalebar{ -0.66 } & -0.66 & [-0.72, -0.61] & *** & \scalebar{ 1.29 } & 1.29 & [1.23, 1.35] & *** \\
News & Reddit & \scalebar{ -0.12 } & -0.12 & [-0.17, -0.08] & *** & \scalebar{ 0.6 } & 0.6 & [0.55, 0.64] & *** \\
\rowcollight Research Pr & News & \scalebar{ 0.46 } & 0.46 & [0.37, 0.55] & *** & \scalebar{ 0.75 } & 0.75 & [0.66, 0.84] & *** \\
\rowcollight Research Pr & YouTube & \scalebar{ -0.48 } & -0.48 & [-0.56, -0.39] & *** & \scalebar{ 1.74 } & 1.74 & [1.65, 1.83] & *** \\
\rowcollight Research Pr & TikTok & \scalebar{ -0.26 } & -0.26 & [-0.34, -0.17] & *** & \scalebar{ 1.45 } & 1.45 & [1.36, 1.54] & *** \\
\rowcollight Research Pr & Reddit & \scalebar{ 0.22 } & 0.22 & [0.14, 0.30] & *** & \scalebar{ 0.85 } & 0.85 & [0.77, 0.93] & *** \\
YouTube & News & \scalebar{ 0.96 } & 0.96 & [0.91, 1.02] & *** & \scalebar{ -1.58 } & -1.58 & [-1.64, -1.52] & *** \\
YouTube & Research Pr & \scalebar{ 0.48 } & 0.48 & [0.39, 0.56] & *** & \scalebar{ -1.74 } & -1.74 & [-1.83, -1.65] & *** \\
YouTube & TikTok & \scalebar{ 0.06 } & 0.06 & [0.01, 0.11] & *** & \scalebar{ -0.16 } & -0.16 & [-0.21, -0.12] & *** \\
YouTube & Reddit & \scalebar{ 0.54 } & 0.54 & [0.50, 0.58] & *** & \scalebar{ -0.63 } & -0.63 & [-0.67, -0.60] & *** \\
\rowcollight TikTok & News & \scalebar{ 0.66 } & 0.66 & [0.61, 0.72] & *** & \scalebar{ -1.29 } & -1.29 & [-1.35, -1.23] & *** \\
\rowcollight TikTok & Research Pr & \scalebar{ 0.26 } & 0.26 & [0.17, 0.34] & *** & \scalebar{ -1.45 } & -1.45 & [-1.54, -1.36] & *** \\
\rowcollight TikTok & YouTube & \scalebar{ -0.06 } & -0.06 & [-0.11, -0.01] & *** & \scalebar{ 0.16 } & 0.16 & [0.12, 0.21] & *** \\
\rowcollight TikTok & Reddit & \scalebar{ 0.46 } & 0.46 & [0.42, 0.50] &  & \scalebar{ -0.47 } & -0.47 & [-0.51, -0.43] & *** \\
Reddit & News & \scalebar{ 0.12 } & 0.12 & [0.08, 0.17] & *** & \scalebar{ -0.6 } & -0.6 & [-0.64, -0.55] & *** \\
Reddit & Research Pr & \scalebar{ -0.22 } & -0.22 & [-0.30, -0.14] & *** & \scalebar{ -0.85 } & -0.85 & [-0.93, -0.77] & *** \\
Reddit & YouTube & \scalebar{ -0.54 } & -0.54 & [-0.58, -0.50] & *** & \scalebar{ 0.63 } & 0.63 & [0.60, 0.67] & *** \\
Reddit & TikTok & \scalebar{ -0.46 } & -0.46 & [-0.50, -0.42] &  & \scalebar{ 0.47 } & 0.47 & [0.43, 0.51] & *** \\
\hline
\\
\multicolumn{2}{l}{} & \multicolumn{4}{c}{\textbf{Clout}} & \multicolumn{4}{c}{\textbf{Authentic}}\\
&  &    & $d$ & 95\% CI & $p$ &    & $d$ & 95\% CI & $p$ \\
\hline
News & Research Pr & \scalebar{ 0.39 } & 0.39 & [0.30, 0.48] & *** & \scalebar{ 0.28 } & 0.28 & [0.19, 0.37] & *** \\
News & YouTube & \scalebar{ -0.49 } & -0.49 & [-0.55, -0.44] & *** & \scalebar{ -0.62 } & -0.62 & [-0.67, -0.57] & *** \\
News & TikTok & \scalebar{ -0.36 } & -0.36 & [-0.42, -0.31] & *** & \scalebar{ -0.46 } & -0.46 & [-0.51, -0.40] & *** \\
News & Reddit & \scalebar{ 0.52 } & 0.52 & [0.47, 0.56] & *** & \scalebar{ -0.49 } & -0.49 & [-0.53, -0.44] & *** \\
\rowcollight Research Pr & News & \scalebar{ -0.39 } & -0.39 & [-0.48, -0.30] & *** & \scalebar{ -0.28 } & -0.28 & [-0.37, -0.19] & *** \\
\rowcollight Research Pr & YouTube & \scalebar{ -0.85 } & -0.85 & [-0.94, -0.76] & *** & \scalebar{ -0.84 } & -0.84 & [-0.93, -0.76] & *** \\
\rowcollight Research Pr & TikTok & \scalebar{ -0.53 } & -0.53 & [-0.62, -0.44] & *** & \scalebar{ -0.56 } & -0.56 & [-0.65, -0.47] &  \\
\rowcollight Research Pr & Reddit & \scalebar{ 0.33 } & 0.33 & [0.25, 0.41] & *** & \scalebar{ -0.58 } & -0.58 & [-0.66, -0.50] & *** \\
YouTube & News & \scalebar{ 0.49 } & 0.49 & [0.44, 0.55] & *** & \scalebar{ 0.62 } & 0.62 & [0.57, 0.67] & *** \\
YouTube & Research Pr & \scalebar{ 0.85 } & 0.85 & [0.76, 0.94] & *** & \scalebar{ 0.84 } & 0.84 & [0.76, 0.93] & *** \\
YouTube & TikTok & \scalebar{ -0.04 } & -0.04 & [-0.09, 0.00] & *** & \scalebar{ -0.03 } & -0.03 & [-0.08, 0.02] & *** \\
YouTube & Reddit & \scalebar{ 0.78 } & 0.78 & [0.74, 0.82] & *** & \scalebar{ -0.2 } & -0.2 & [-0.23, -0.16] & *** \\
\rowcollight TikTok & News & \scalebar{ 0.36 } & 0.36 & [0.31, 0.42] & *** & \scalebar{ 0.46 } & 0.46 & [0.40, 0.51] & *** \\
\rowcollight TikTok & Research Pr & \scalebar{ 0.53 } & 0.53 & [0.44, 0.62] & *** & \scalebar{ 0.56 } & 0.56 & [0.47, 0.65] &  \\
\rowcollight TikTok & YouTube & \scalebar{ 0.04 } & 0.04 & [-0.00, 0.09] & *** & \scalebar{ 0.03 } & 0.03 & [-0.02, 0.08] & *** \\
\rowcollight TikTok & Reddit & \scalebar{ 0.76 } & 0.76 & [0.72, 0.80] & * & \scalebar{ -0.17 } & -0.17 & [-0.20, -0.13] & *** \\
Reddit & News & \scalebar{ -0.52 } & -0.52 & [-0.56, -0.47] & *** & \scalebar{ 0.49 } & 0.49 & [0.44, 0.53] & *** \\
Reddit & Research Pr & \scalebar{ -0.33 } & -0.33 & [-0.41, -0.25] & *** & \scalebar{ 0.58 } & 0.58 & [0.50, 0.66] & *** \\
Reddit & YouTube & \scalebar{ -0.78 } & -0.78 & [-0.82, -0.74] & *** & \scalebar{ 0.2 } & 0.2 & [0.16, 0.23] & *** \\
Reddit & TikTok & \scalebar{ -0.76 } & -0.76 & [-0.80, -0.72] & * & \scalebar{ 0.17 } & 0.17 & [0.13, 0.20] & *** \\
\bottomrule
\end{tabular}
\caption{Platform pairwise comparison for lexical styles. Cohen's d with Dunn's test p-values (Bonferroni corrected) where *** $p<0.001$, ** $p<0.01$, * $p<0.05$. Across platforms, YouTube showed a significantly more positive tone, Research Press showed significantly higher analyticity, TikTok showed significantly greater clout, and Reddit showed significantly higher authenticity.}
\label{table:pairwise_style}
\end{table*}

\begin{table*}[]
\centering
\sffamily
\footnotesize
\begin{tabular}{llrrclrrclrrcl}
\toprule
\multicolumn{2}{l}{} & \multicolumn{4}{c}{\textbf{Risks to individuals}} & \multicolumn{4}{c}{\textbf{Risks to human-centered care}} & \multicolumn{4}{c}{\textbf{Risks to information ecosystems}}\\
 &  &   & $d$ & 95\% CI & $p$ &   & $d$ & 95\% CI & $p$ &   & $d$ & 95\% CI & $p$ \\
\hline
News & Research Pr & \scalebar{ 0.4 } & 0.4 & [0.31, 0.49] & *** & \scalebar{ 0.23 } & 0.23 & [0.15, 0.32] & *** & \scalebar{ 0.52 } & 0.52 & [0.43, 0.61] & *** \\
News & YouTube & \scalebar{ -0.01 } & -0.01 & [-0.06, 0.04] & *** & \scalebar{ 0.18 } & 0.18 & [0.12, 0.23] & *** & \scalebar{ 0.07 } & 0.07 & [0.01, 0.12] & *** \\
News & TikTok & \scalebar{ 0.91 } & 0.91 & [0.86, 0.97] & *** & \scalebar{ 0.76 } & 0.76 & [0.70, 0.82] & *** & \scalebar{ 1.98 } & 1.98 & [1.91, 2.04] & *** \\
News & Reddit & \scalebar{ 0.47 } & 0.47 & [0.43, 0.52] &  & \scalebar{ 0.75 } & 0.75 & [0.70, 0.79] & *** & \scalebar{ 1.06 } & 1.06 & [1.02, 1.11] &  \\
\rowcollight Research Pr & News & \scalebar{ -0.4 } & -0.4 & [-0.49, -0.31] & *** & \scalebar{ -0.23 } & -0.23 & [-0.32, -0.15] & *** & \scalebar{ -0.52 } & -0.52 & [-0.61, -0.43] & *** \\
\rowcollight Research Pr & YouTube & \scalebar{ -0.41 } & -0.41 & [-0.49, -0.32] &  & \scalebar{ -0.07 } & -0.07 & [-0.15, 0.02] & *** & \scalebar{ -0.44 } & -0.44 & [-0.53, -0.36] & *** \\
\rowcollight Research Pr & TikTok & \scalebar{ 0.49 } & 0.49 & [0.41, 0.58] & *** & \scalebar{ 0.62 } & 0.62 & [0.54, 0.71] & *** & \scalebar{ 1.37 } & 1.37 & [1.28, 1.46] & *** \\
\rowcollight Research Pr & Reddit & \scalebar{ 0.07 } & 0.07 & [-0.01, 0.15] & *** & \scalebar{ 0.44 } & 0.44 & [0.36, 0.52] & *** & \scalebar{ 0.56 } & 0.56 & [0.48, 0.64] & *** \\
YouTube & News & \scalebar{ 0.01 } & 0.01 & [-0.04, 0.06] & *** & \scalebar{ -0.18 } & -0.18 & [-0.23, -0.12] & *** & \scalebar{ -0.07 } & -0.07 & [-0.12, -0.01] & *** \\
YouTube & Research Pr & \scalebar{ 0.41 } & 0.41 & [0.32, 0.49] &  & \scalebar{ 0.07 } & 0.07 & [-0.02, 0.15] & *** & \scalebar{ 0.44 } & 0.44 & [0.36, 0.53] & *** \\
YouTube & TikTok & \scalebar{ 0.9 } & 0.9 & [0.85, 0.95] & *** & \scalebar{ 0.53 } & 0.53 & [0.48, 0.58] & *** & \scalebar{ 1.78 } & 1.78 & [1.72, 1.84] & *** \\
YouTube & Reddit & \scalebar{ 0.48 } & 0.48 & [0.44, 0.52] & *** & \scalebar{ 0.48 } & 0.48 & [0.44, 0.52] &  & \scalebar{ 1 } & 1 & [0.96, 1.04] & *** \\
\rowcollight TikTok & News & \scalebar{ -0.91 } & -0.91 & [-0.97, -0.86] & *** & \scalebar{ -0.76 } & -0.76 & [-0.82, -0.70] & *** & \scalebar{ -1.98 } & -1.98 & [-2.04, -1.91] & *** \\
\rowcollight TikTok & Research Pr & \scalebar{ -0.49 } & -0.49 & [-0.58, -0.41] & *** & \scalebar{ -0.62 } & -0.62 & [-0.71, -0.54] & *** & \scalebar{ -1.37 } & -1.37 & [-1.46, -1.28] & *** \\
\rowcollight TikTok & YouTube & \scalebar{ -0.9 } & -0.9 & [-0.95, -0.85] & *** & \scalebar{ -0.53 } & -0.53 & [-0.58, -0.48] & *** & \scalebar{ -1.78 } & -1.78 & [-1.84, -1.72] & *** \\
\rowcollight TikTok & Reddit & \scalebar{ -0.37 } & -0.37 & [-0.41, -0.33] & *** & \scalebar{ -0.13 } & -0.13 & [-0.16, -0.09] & *** & \scalebar{ -0.48 } & -0.48 & [-0.52, -0.44] & *** \\
Reddit & News & \scalebar{ -0.47 } & -0.47 & [-0.52, -0.43] &  & \scalebar{ -0.75 } & -0.75 & [-0.79, -0.70] & *** & \scalebar{ -1.06 } & -1.06 & [-1.11, -1.02] &  \\
Reddit & Research Pr & \scalebar{ -0.07 } & -0.07 & [-0.15, 0.01] & *** & \scalebar{ -0.44 } & -0.44 & [-0.52, -0.36] & *** & \scalebar{ -0.56 } & -0.56 & [-0.64, -0.48] & *** \\
Reddit & YouTube & \scalebar{ -0.48 } & -0.48 & [-0.52, -0.44] & *** & \scalebar{ -0.48 } & -0.48 & [-0.52, -0.44] &  & \scalebar{ -1 } & -1 & [-1.04, -0.96] & *** \\
Reddit & TikTok & \scalebar{ 0.37 } & 0.37 & [0.33, 0.41] & *** & \scalebar{ 0.13 } & 0.13 & [0.09, 0.16] & *** & \scalebar{ 0.48 } & 0.48 & [0.44, 0.52] & *** \\
\hline
\\
\multicolumn{2}{l}{} & \multicolumn{4}{c}{\textbf{Risks to technology accountability}} & \multicolumn{4}{c}{\textbf{}} & \multicolumn{4}{c}{\textbf{Explanations of generative nature}}\\
&  &   & $d$ & 95\% CI & $p$ &   &  &  & &   & $d$ & 95\% CI & $p$ \\
\hline
News &Research Pr &\scalebar{ 0.41 }&0.41 &[0.32, 0.50] &*** && & & &\scalebar{ 0.03 }&0.03 &[-0.06, 0.11]&** \\
News &YouTube &\scalebar{ 0.03 }&0.03 &[-0.02, 0.09]&*** && & & &\scalebar{ -0.74}&-0.74&[-0.80, -0.69] & \\
News &TikTok &\scalebar{ 1.23 }&1.23 &[1.17, 1.29] &*** && & & &\scalebar{ 0.25 }&0.25 &[0.19, 0.30] &*** \\
News &Reddit &\scalebar{ 0.89 }&0.89 &[0.84, 0.94] & && & & &\scalebar{ 0.16 }&0.16 &[0.11, 0.20] &*** \\
\rowcollight Research Pr &News &\scalebar{ -0.41}&-0.41&[-0.50, -0.32] &*** && & & &\scalebar{ -0.03}&-0.03&[-0.11, 0.06]&** \\
\rowcollight Research Pr &YouTube &\scalebar{ -0.37}&-0.37&[-0.46, -0.29] &*** && & & &\scalebar{ -0.66}&-0.66&[-0.75, -0.58] & \\
\rowcollight Research Pr &TikTok &\scalebar{ 0.82 }&0.82 &[0.74, 0.91] &*** && & & &\scalebar{ 0.31 }&0.31 &[0.23, 0.40] & \\
\rowcollight Research Pr &Reddit &\scalebar{ 0.43 }&0.43 &[0.35, 0.51] &*** && & & &\scalebar{ 0.13 }&0.13 &[0.05, 0.21] &*** \\
YouTube &News &\scalebar{ -0.03}&-0.03&[-0.09, 0.02]&*** && & & &\scalebar{ 0.74 }&0.74 &[0.69, 0.80] & \\
YouTube &Research Pr &\scalebar{ 0.37 }&0.37 &[0.29, 0.46] &*** && & & &\scalebar{ 0.66 }&0.66 &[0.58, 0.75] & \\
YouTube &TikTok &\scalebar{ 1.13 }&1.13 &[1.08, 1.18] &*** && & & &\scalebar{ 0.92 }&0.92 &[0.87, 0.97] & \\
YouTube &Reddit &\scalebar{ 0.84 }&0.84 &[0.80, 0.88] &*** && & & &\scalebar{ 1.25 }&1.25 &[1.21, 1.29] &*** \\
\rowcollight TikTok &News &\scalebar{ -1.23}&-1.23&[-1.29, -1.17] &*** && & & &\scalebar{ -0.25}&-0.25&[-0.30, -0.19] &*** \\
\rowcollight TikTok &Research Pr &\scalebar{ -0.82}&-0.82&[-0.91, -0.74] &*** && & & &\scalebar{ -0.31}&-0.31&[-0.40, -0.23] & \\
\rowcollight TikTok &YouTube &\scalebar{ -1.13}&-1.13&[-1.18, -1.08] &*** && & & &\scalebar{ -0.92}&-0.92&[-0.97, -0.87] & \\
\rowcollight TikTok &Reddit &\scalebar{ -0.28}&-0.28&[-0.31, -0.24] &*** && & & &\scalebar{ -0.1 }&-0.1 &[-0.14, -0.06] &*** \\
Reddit &News &\scalebar{ -0.89}&-0.89&[-0.94, -0.84] & && & & &\scalebar{ -0.16}&-0.16&[-0.20, -0.11] &*** \\
Reddit &Research Pr &\scalebar{ -0.43}&-0.43&[-0.51, -0.35] &*** && & & &\scalebar{ -0.13}&-0.13&[-0.21, -0.05] &*** \\
Reddit &YouTube &\scalebar{ -0.84}&-0.84&[-0.88, -0.80] &*** && & & &\scalebar{ -1.25}&-1.25&[-1.29, -1.21] &*** \\
Reddit &TikTok &\scalebar{ 0.28 }&0.28 &[0.24, 0.31] &*** && & & &\scalebar{ 0.1 }&0.1 &[0.06, 0.14] &*** \\
\bottomrule
\end{tabular}
\caption{Platform pairwise comparison for informational content: disclosure of risks and generative nature. Cohen's d with Dunn's test p-values (Bonferroni corrected) where *** $p<0.001$, ** $p<0.01$, * $p<0.05$.}
\label{table:pairwise_risk}
\end{table*}

\begin{table*}[]
\centering
\sffamily
\footnotesize
\begin{tabular}{llrrcl}
\toprule
\multicolumn{2}{l}{} & \multicolumn{4}{c}{\textbf{Anthropomorphism}}\\
&  &   & $d$ & 95\% CI & $p$ \\
\hline
News & Research Pr & \scalebar{ 3.21 } & 3.21 & [3.09, 3.33] & *** \\
News & YouTube & \scalebar{ 1 } & 1 & [0.94, 1.05] & *** \\
News & TikTok & \scalebar{ -0.28 } & -0.28 & [-0.33, -0.22] & *** \\
News & Reddit & \scalebar{ -0.28 } & -0.28 & [-0.32, -0.23] & *** \\
\rowcollight Research Pr & News & \scalebar{ -3.21 } & -3.21 & [-3.33, -3.09] & *** \\
\rowcollight Research Pr & YouTube & \scalebar{ -3.18 } & -3.18 & [-3.29, -3.07] & *** \\
\rowcollight Research Pr & TikTok & \scalebar{ -1.85 } & -1.85 & [-1.95, -1.75] &  \\
\rowcollight Research Pr & Reddit & \scalebar{ -2.43 } & -2.43 & [-2.52, -2.35] & *** \\
YouTube & News & \scalebar{ -1 } & -1 & [-1.05, -0.94] & *** \\
YouTube & Research Pr & \scalebar{ 3.18 } & 3.18 & [3.07, 3.29] & *** \\
YouTube & TikTok & \scalebar{ -0.77 } & -0.77 & [-0.82, -0.72] & *** \\
YouTube & Reddit & \scalebar{ -0.8 } & -0.8 & [-0.84, -0.76] & *** \\
\rowcollight TikTok & News & \scalebar{ 0.28 } & 0.28 & [0.22, 0.33] & *** \\
\rowcollight TikTok & Research Pr & \scalebar{ 1.85 } & 1.85 & [1.75, 1.95] &  \\
\rowcollight TikTok & YouTube & \scalebar{ 0.77 } & 0.77 & [0.72, 0.82] & *** \\
\rowcollight TikTok & Reddit & \scalebar{ 0.04 } & 0.04 & [0.01, 0.08] & *** \\
Reddit & News & \scalebar{ 0.28 } & 0.28 & [0.23, 0.32] & *** \\
Reddit & Research Pr & \scalebar{ 2.43 } & 2.43 & [2.35, 2.52] & *** \\
Reddit & YouTube & \scalebar{ 0.8 } & 0.8 & [0.76, 0.84] & *** \\
Reddit & TikTok & \scalebar{ -0.04 } & -0.04 & [-0.08, -0.01] & *** \\
\bottomrule
\end{tabular}
\caption{Platform pairwise comparison for symbolic representation. Cohen's d with Dunn's test p-values (Bonferroni corrected) where *** $p<0.001$, ** $p<0.01$, * $p<0.05$.}
\label{table:pairwise_sumbolic}
\end{table*}

\begin{figure*}
\centering
\sffamily
\label{fig:trends_all}
\section{Temporal Trends}
\label{appendix:trends}
    \begin{subfigure}[]{\columnwidth}
        \includegraphics[width=\columnwidth]{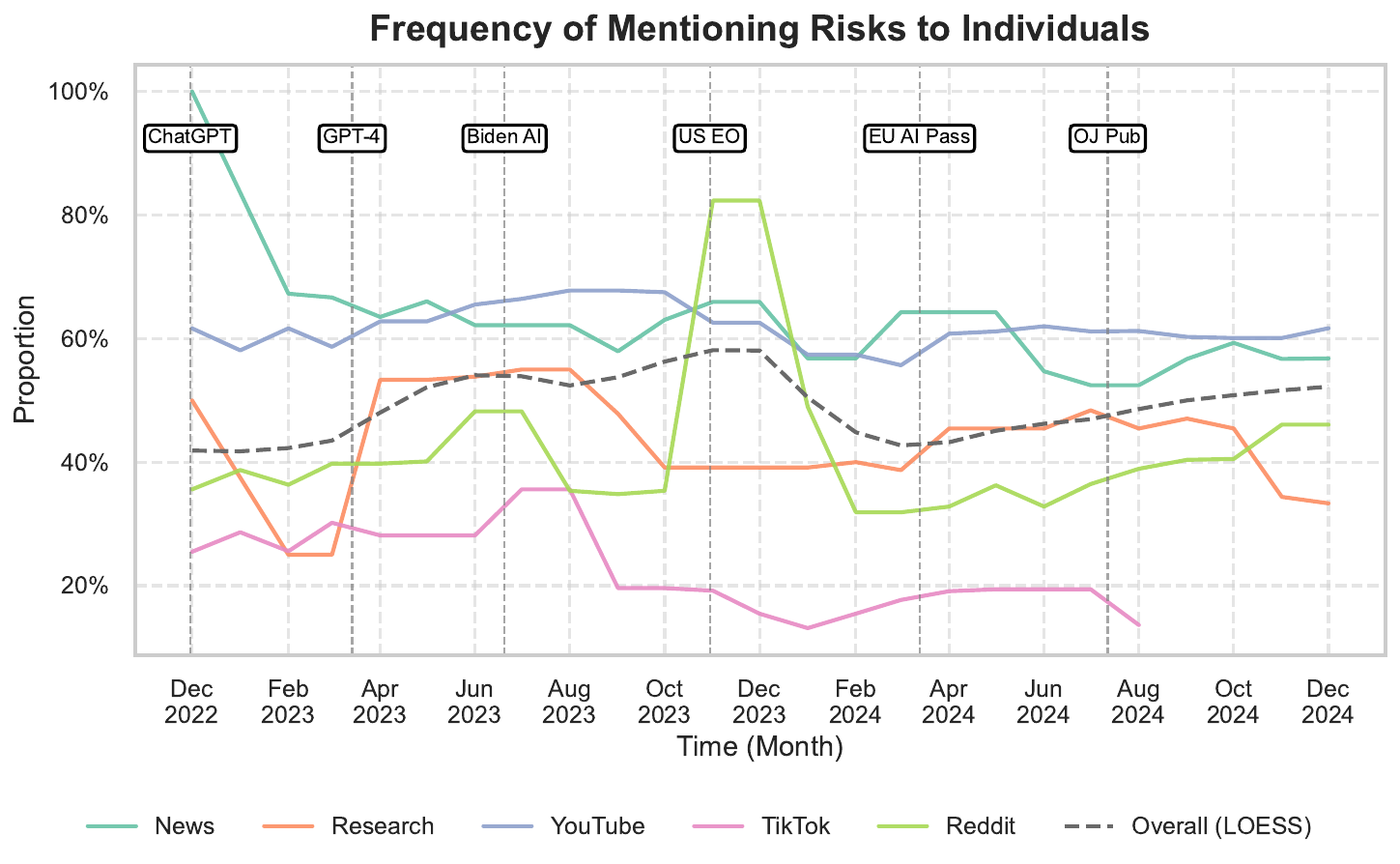}
        \caption{Mentions of risks to individuals exhibited a slight increase over time, while TikTok showed a decline in disclosure frequency.}
    \end{subfigure}
    \hfill
    \begin{subfigure}[]{\columnwidth}
        \includegraphics[width=\columnwidth]{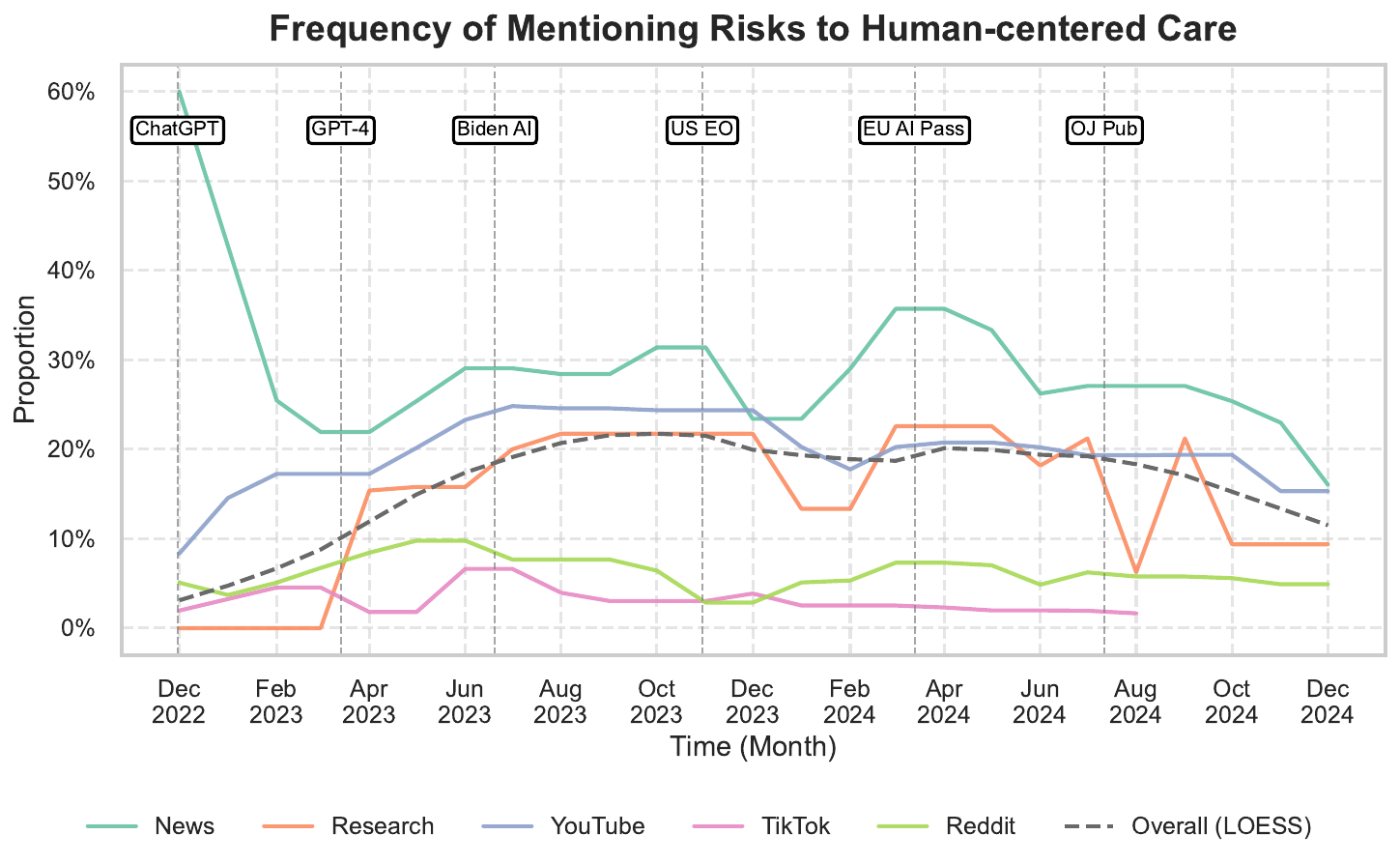}
        \caption{Mentions of risks to human-centered care increased gradually but remained secondary to other risk categories.}
    \end{subfigure}
    \hfill
    \begin{subfigure}[]{\columnwidth}
        \centering
        \includegraphics[width=\columnwidth]{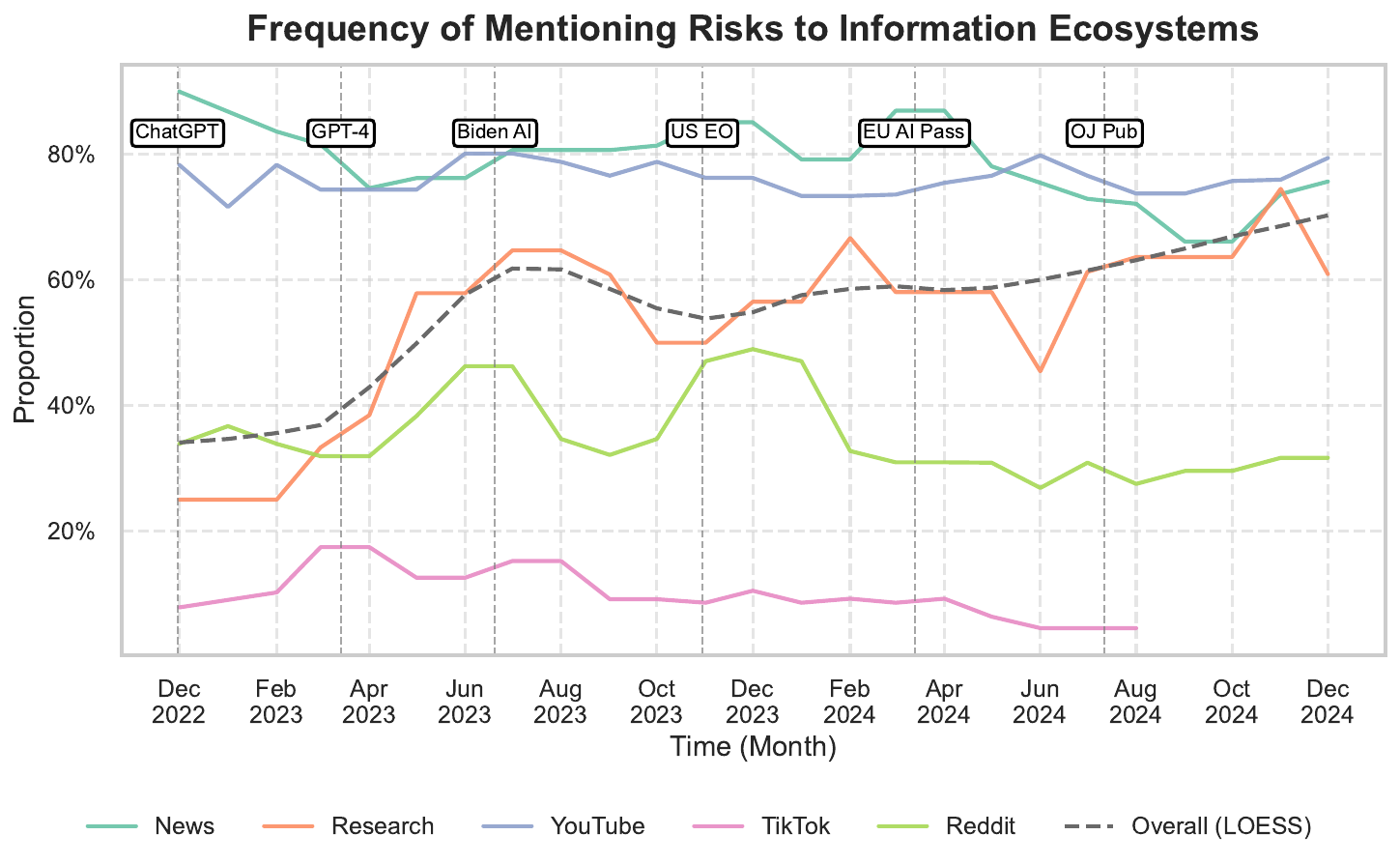}
        \caption{Overall, mentions of risks to information ecosystems were most frequent and grew steadily over time, but TikTok and Reddit had similar or lower frequencies of discussing related risks.}
    \end{subfigure}
    \hfill
    \begin{subfigure}[]{\columnwidth}
        \centering
        \includegraphics[width=\columnwidth]{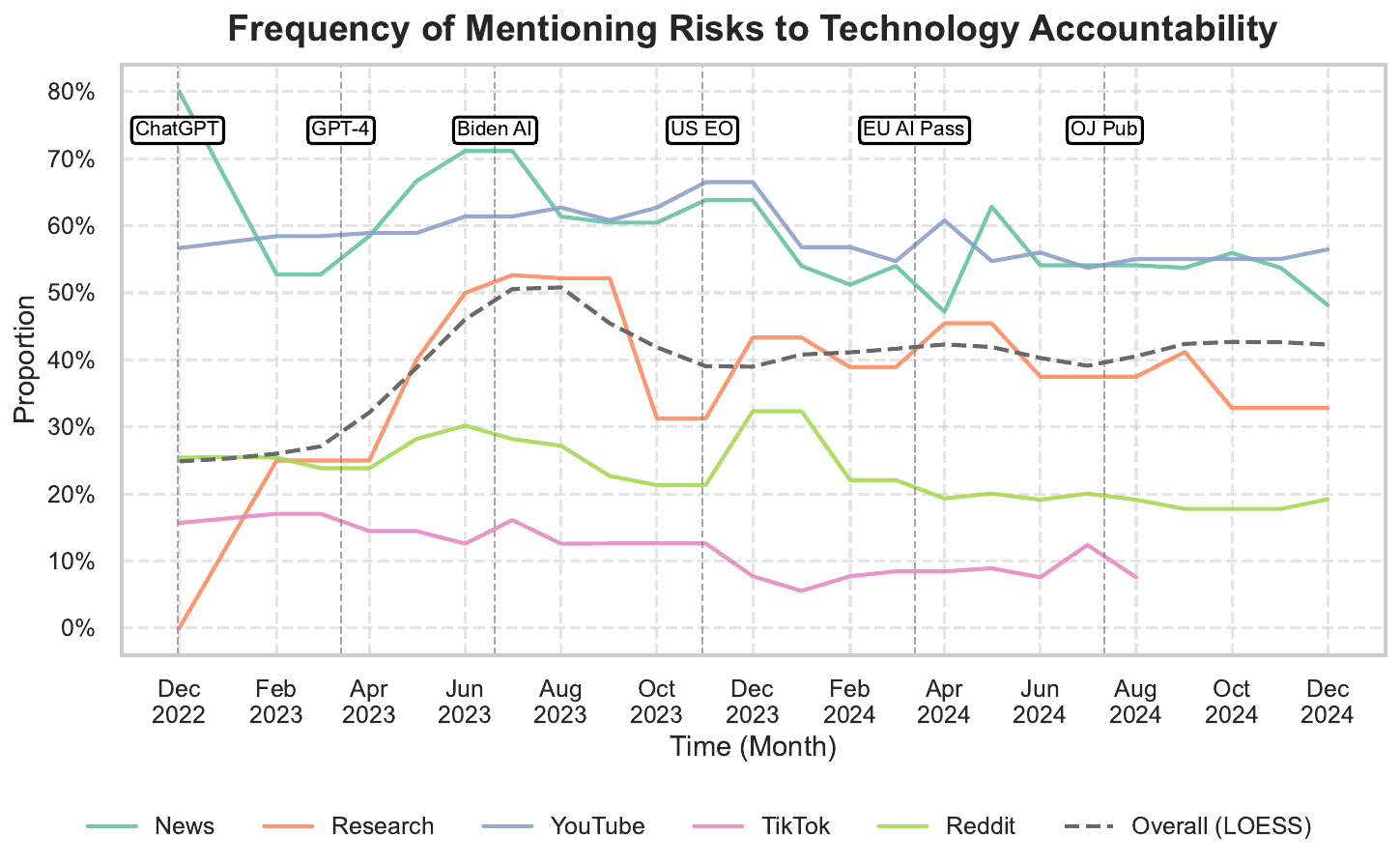}
        \caption{Mentions of risks to technology accountability increased over time, with news and research press being more likely to do so.}
    \end{subfigure}\hfill
    \begin{subfigure}[]{\columnwidth}
        \centering
        \includegraphics[width=\columnwidth]{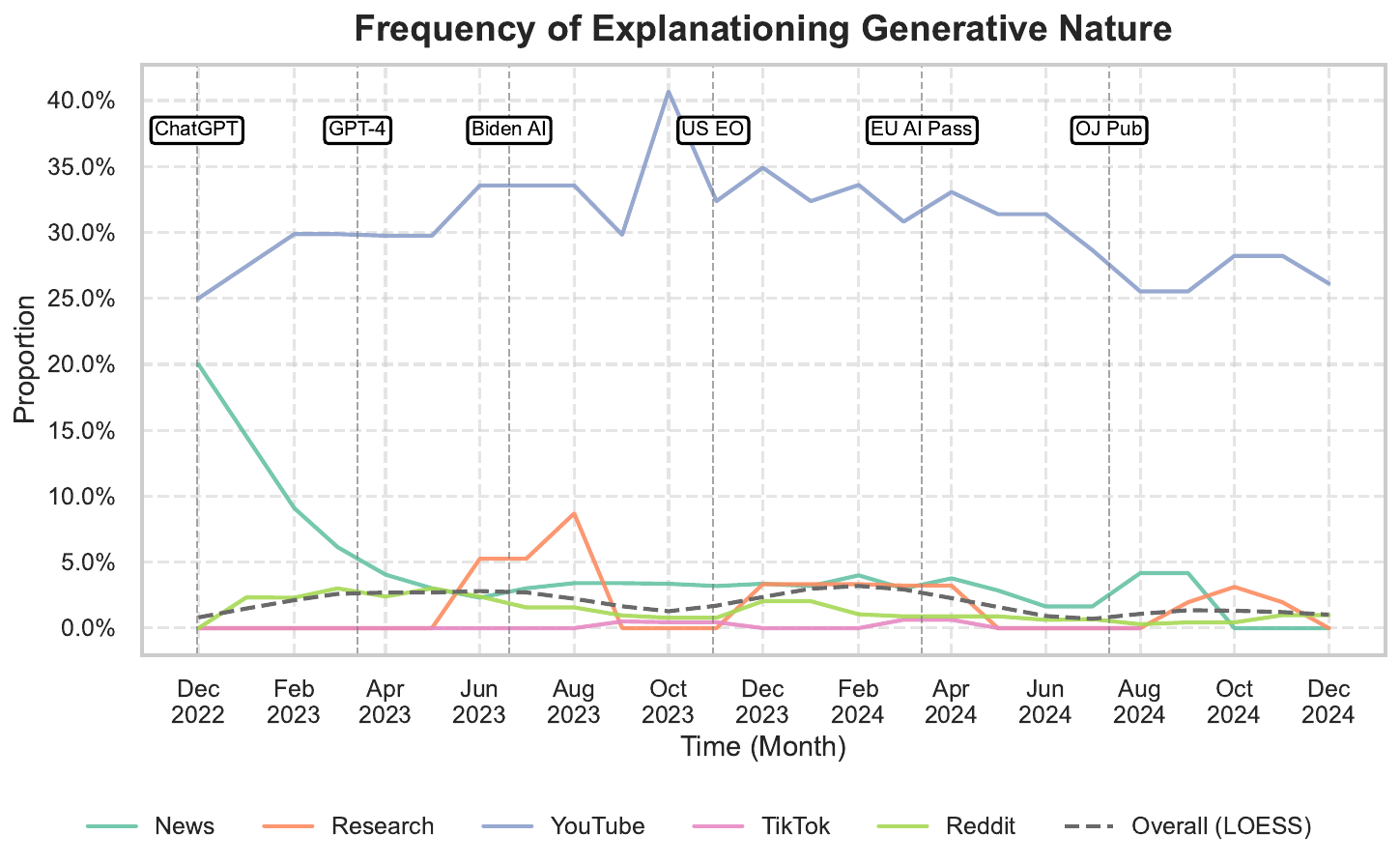}
        \caption{Across platforms, explicit explanations of LLMs' generative nature remained extremely low over time.}
    \end{subfigure}
    \hfill
    \begin{subfigure}[]{\columnwidth}
        \centering
        \includegraphics[width=\columnwidth]{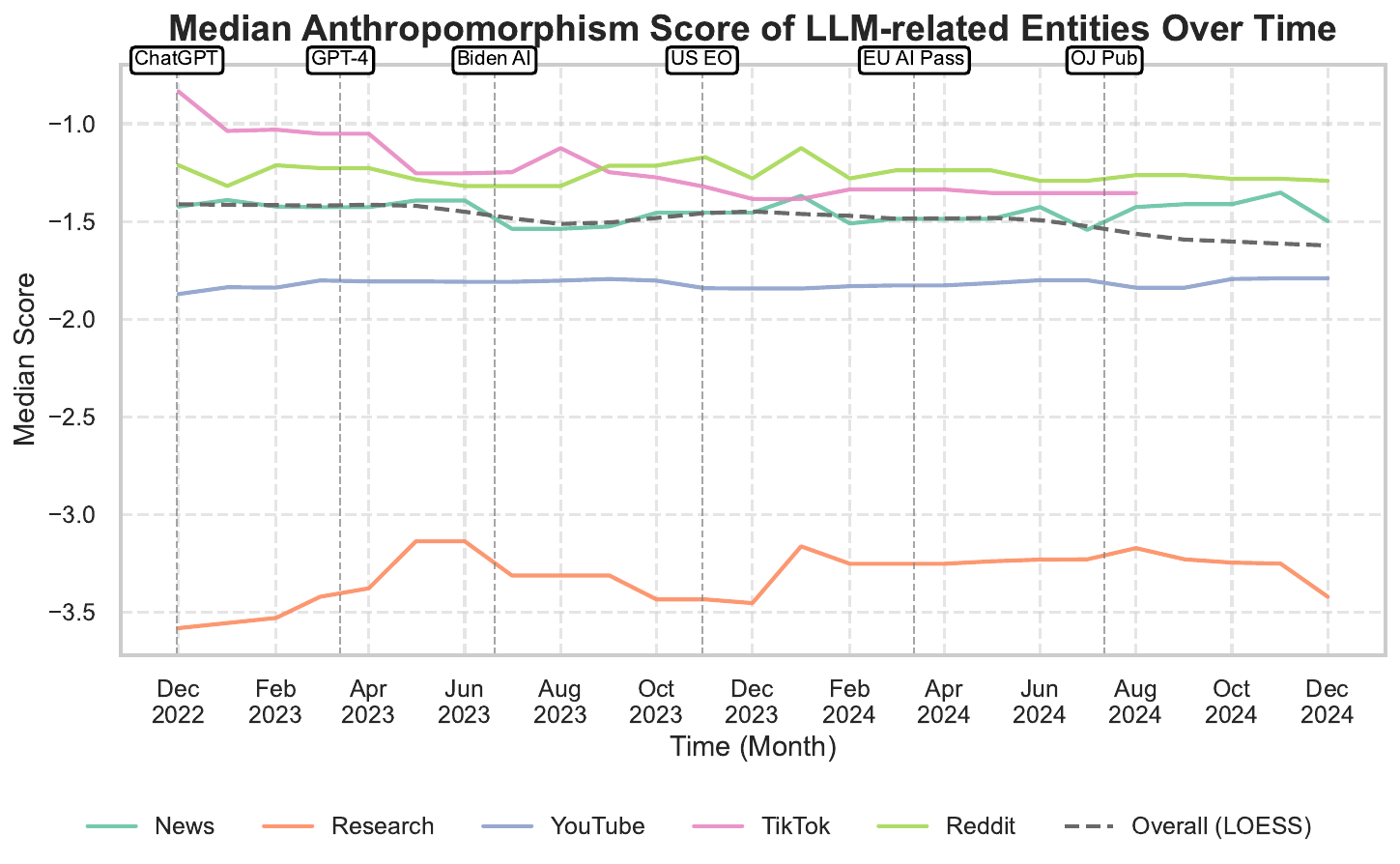}
        \caption{The median anthropomorphic representation of LLMs declined slightly over time but showed a substantial divergence between research press and other discourse channels.}
    \end{subfigure}
\caption{Temporal trends. TikTok data ended in August 2024 due to an unresolved API internal error (see Sec.~\ref{section:data_collection}). We applied LOESS (Locally Estimated Scatterplot Smoothing) to capture overall trends across five data sources of varying size and nature. }
\end{figure*}

\end{document}